\documentclass[11pt,a4paper]{article}
\pdfoutput=1
\usepackage{jheppub}
\usepackage{amsfonts,amsthm}
\usepackage{amsmath,amssymb}
\usepackage{graphicx,color}
\usepackage{multirow}

\newcommand{\be}{\begin{equation}}
\newcommand{\ee}{\end{equation}}
\newcommand{\bea}{\begin{eqnarray}}
\newcommand{\eea}{\end{eqnarray}}

\renewcommand{\Re}{\mathrm{Re}\,}
\renewcommand{\Im}{\mathrm{Im}\,}
\newcommand{\doublet}[2]{ \left( \begin{array}{c}#1 \\ #2 \end{array}\right) }
\newcommand{\triplet}[3]{ \left(\! \begin{array}{c} {#1} \\ {#2}  \\ {#3}\end{array}\!\right) }

\newcommand{\lr}[1]{ \langle #1 \rangle}
\newcommand{\Tr}{\mathrm{Tr}}
\newcommand{\Z}{\mathbb{Z}}
\newcommand{\mmatrix}[4]{ \left(\! \begin{array}{ccc}#1 & #2 \\ #3 & #4 \end{array}\!\right) }
\newcommand{\mmmatrix}[9]{ \left(\! \begin{array}{ccc}#1 & #2 & #3\\ #4 & #5 & #6\\ #7 & #8 & #9\\ \end{array}\!\right) }
\newcommand{\toCP}{\xrightarrow{CP}}

\newcommand{\bx}{{\bf x}}

\def\lsim{\mathrel{\rlap{\lower4pt\hbox{\hskip1pt$\sim$}}
    \raise1pt\hbox{$<$}}}         
\def\gsim{\mathrel{\rlap{\lower4pt\hbox{\hskip1pt$\sim$}}
    \raise1pt\hbox{$>$}}}         

\preprint{LU TP 17-35}

\title{CP4 miracle: shaping Yukawa sector with CP symmetry of order four}

\author[a]{P.M. Ferreira,}
\author[b]{Igor P. Ivanov,}
\author[c,d]{Enrique Jim\'enez,}
\author[e]{Roman Pasechnik,}
\author[e]{Hugo Ser\^{o}dio}

\affiliation[a] {​Centro de Física Te\'orica e Computacional --- FCUL, Universidade de Lisboa,
R. Ernesto de Vasconcelos, 1749--016 Lisboa, Portugal}
\affiliation[b] {CFTP, Instituto Superior T\'ecnico, Universidade de Lisboa, Avenida Rovisco Pais 1, 1049--001 Lisboa, Portugal}
\affiliation[c] {Facultad de Ciencias --- CUICBAS, Universidad de Colima, C.P. 28045, Colima, M\'exico}
\affiliation[d] {Dual CP Institute of High Energy Physics, C.P. 28045, Colima, M\'exico}
\affiliation[e] {Department of Astronomy and Theoretical Physics, Lund University, SE-223 62 Lund, Sweden}

\emailAdd{pmmferreira@fc.ul.pt}
\emailAdd{igor.ivanov@tecnico.ulisboa.pt}
\emailAdd{physieira@gmail.com}
\emailAdd{roman.pasechnik@thep.lu.se}
\emailAdd{hugo.serodio@thep.lu.se}

\date{\today}

\abstract{
We explore the phenomenology of a unique three-Higgs-doublet model based on the single $CP$ symmetry
of order 4 (CP4) without any accidental symmetries. The CP4 symmetry is imposed on the scalar potential and Yukawa interactions,
strongly shaping both sectors of the model and leading to a very characteristic phenomenology. The scalar sector is analyzed
in detail, and in the Yukawa sector we list all possible CP4-symmetric structures which do not run into immediate conflict with
experiment, namely, do not lead to massless or mass-degenerate quarks nor to insufficient mixing or $CP$-violation in
the CKM matrix. 
We show that the parameter space of the model, although very constrained by CP4, is large enough to comply with
the electroweak precision data and the LHC results for the 125 GeV Higgs boson phenomenology, as well as to perfectly reproduce
all fermion masses, mixing, and $CP$ violation. Despite the presence of flavor changing neutral currents mediated by heavy Higgs
scalars, we find through a parameter space scan many points which accurately reproduce the kaon $CP$-violating parameter
$\epsilon_K$ as well as oscillation parameters in $K$ and $B_{(s)}$ mesons. Thus, CP4 offers a novel minimalistic framework
for building models with very few assumptions, sufficient predictive power, and rich phenomenology yet to be explored.
}
\begin{document}
\maketitle

\section{Introduction}

The discovery of the Higgs boson~\cite{:2012gk,:2012gu} was a major milestone in particle physics:
the last remaining piece of the Standard Model (SM) was brought to light. From this point onwards,
any deviations from the observed SM-like character of particle physics could well be a sign of new physics
hitherto undiscovered. And although further LHC measurements have shown that the discovered 125 GeV
scalar is conforming to the SM behaviour (see Ref.~\cite{Khachatryan:2016vau} for the combined
LHC Run I results; and the presentations of the latest Run II results at the recent EPS-HEP 2017 conference
in Venice), the SM leaves too many phenomena unexplained to be regarded as a satisfactory theory.
These include: the absence of dark matter (DM) candidates; complete ignorance of the origin of neutrino
masses \cite{King:2017guk} and explanation of their observed mixings; inability to explain the origin
of the observed $CP$ violation (CPV) \cite{book}; and incapacity to cast any light on the quark and lepton
mass and mixing hierarchies.

Addressing these unsolved questions requires going beyond the Standard Model (BSM), and many proposed
solutions involve expanding the minimal scalar sector of the SM by adding extra scalar fields; for a recent overview
of models with extended Higgs sectors, see Ref.~\cite{Ivanov:2017dad}. A direction, which is particularly
attractive due to its conceptual simplicity, is to stay with Higgs $SU(2)$ doublets and to extend the notion of generations to the scalar sector.
One arrives in this way to $N$-Higgs-doublet models (NHDM). The Two-Higgs Doublet Model (2HDM) (see Ref.~\cite{2HDM-review} for a review)
is the most popular example but models with more Higgs doublets are also being actively explored. Such extended scalar sectors have a rich
phenomenology but they come with a price: as the number of extra fields grow, so does the number of free parameters, and thus the predictive
power of the theory is reduced. When building such models, it would be desirable to achieve a balance between the following two requirements:
making as few extra assumptions as possible, on top of those in the SM, and obtaining a model
which satisfies all experimental constraints, while also being able to make testable predictions
for the ongoing or future measurements. One would like to avoid describing all available data
at the cost of an excessive number of new fields and parameters, but also to avoid obtaining
a neat BSM model, so tightly constrained by theoretical constraints that it fails a comparison
with the experimental bounds.

A successful way of reducing the number of free parameters of BSM models, thus increasing their
predictivity and even resolving some of their theoretical problems, is by imposing additional
global symmetries, either continuous or discrete \cite{Ishimori:2010au,King:2017guk}. For instance,
the most general 2HDM has Higgs-mediated flavour-changing neutral currents (FCNC), but by
enforcing its Lagrangian to be invariant under a discrete $Z_2$ symmetry, those undesirable
interactions are made to vanish~\cite{Glashow:1976nt,Paschos:1976ay,Peccei:1977hh}.
Generically, a typical $N$-Higgs-Doublet Model (NHDM) will contain hundreds of free parameters in its scalar
and Yukawa sectors. By imposing large non-Abelian discrete symmetry groups this number of parameters may be reduced
to about a dozen, making such a model rather predictive. However, even though it is reasonably easy to fashion models
with an acceptable scalar sector --- {\em i.e.}, models which include a scalar state of mass 125 GeV
with a SM-like behaviour and extra scalars that are not yet excluded by the LHC searches --- such models usually
render fermionic sectors which are unphysical \cite{Felipe:2014zka}. Indeed, for sufficiently large discrete symmetry groups,
there is always some residual symmetry preserved by the vacuum, which will imply either massless or
mass-degenerate fermions, or alternatively, lead to an inadequate quark mixing, or an insufficient CPV.
Smaller symmetry groups may lead to good experimental fits, but they usually leave the model with an excess of free
parameters, making such theories cumbersome to analyse and less attractive as an alternative to the SM.

Recently, in Refs.~\cite{Ivanov:2015mwl,Aranda:2016qmp} a new type of a multi-Higgs model was proposed --- based on
a single symmetry requirement which, rather surprisingly, leads to well-shaped scalar and fermion sectors. The symmetry
assumption behind this model is very simple:
\begin{center}
\parbox{12cm}{
The minimal multi-Higgs-doublet model implementing a $CP$-symmetry of higher order
without producing any accidental symmetry.}
\end{center}
Two aspects here require clarification. First, the order $k$ of a symmetry transformation is the minimal number of times
one needs to apply it to arrive at the identity transformation. The traditionally defined $CP$-transformation is of order 2,
but higher-order (generalized) $CP$-symmetries can also be defined and used for model building, see Section~\ref{section:CP-freedom}.
Second, the concept of {\em accidental symmetries}, which is familiar to model builders in BSM physics, refers to the situation when
requiring that a given model is invariant under a certain type of symmetry produces a Lagrangian which is, in fact, invariant under
larger symmetry groups. Appearance of accidental symmetries signals a certain lack of control over the symmetry content
of a model or its phenomenological manifestations.

The above theoretical request leads to a Three-Higgs Doublet Model (3HDM) with an order-4 $CP$-symmetry, which we label CP4.
This model has a scalar potential which was initially written in Ref.~\cite{abelian} and explored, in the case of
unbroken CP4, in Ref.~\cite{Ivanov:2015mwl}.
If CP4 stays intact, the model incorporates a novel feature in BSM models: a complex scalar field which, although being
a $CP$-eigenstate, is neither $CP$-even nor $CP$-odd but rather, in this case, a $CP$-half-odd state.
In Ref.~\cite{Aranda:2016qmp}, it was shown that the CP4 symmetry can be extended to the Yukawa sector
in a satisfactory manner. However, in order to avoid unobserved mass-degenerate fermions, the CP4 symmetry
must be spontaneously broken. The question of whether the CP4 3HDM Lagrangian can fit the current experimental
data was left unanswered up until now, and that is the main objective of this work.

In short, the CP4 3HDM emerges as an interesting candidate of a model which keeps a fair balance between the minimality
of its theoretical assumptions, and its phenomenological richness and predictivity. We firmly believe that a deeper study
of this model will reveal hidden features which are not present in other models.

In this paper, we will undertake the first detailed phenomenological exploration of the CP4 3HDM with the spontaneously
broken CP4 symmetry. Our main objectives will be:
\begin{itemize}
\item
To prove that the model has a scalar sector which complies with known experimental results, 
at least for the SM-like scalar state of mass 125 GeV.
\item
To show that the Yukawa sector of the model, after spontaneous breaking of CP4, can fit the fermion masses, mixing,
and observed $CP$-violation quantities.
\item
To analyse the inevitable tree-level FCNCs of the model and to show that the model has regions of the parameter space
for which they are under control and conform to the experimental measurements.
\end{itemize}
In Section~\ref{section:scalar-sector}, we will discuss the scalar sector of the CP4 3HDM, including the extremization conditions,
the scalar mass matrices, and the conditions for an alignment limit in the scalar sector, which will produce a SM-like
scalar state keeping its FCNCs under control. In Section~\ref{section:yukawa-sector}, we will construct all CP4-symmetric
Yukawa sectors which do not run into an immediate conflict with experiment. We then present in Section~\ref{section:scan}
the results of a dedicated and efficient scan of the scalar and Yukawa parameter space of the model, showing that there exist
 parameter space regions for which the most stringent experimental constraints on FCNCs are satisfied. We close this study with
a discussion and conclusions in Section~\ref{section:discussion}. Several Appendices provide supplementary details.


\section{The scalar sector of CP4 3HDM}
\label{section:scalar-sector}

We will be working with three $SU(2)$ doublets (i.e. 3HDM), each with hypercharge $Y = 1$ and denoted as $\phi_1$, $\phi_2$
and $\phi_3$. {\em A priori}, the scalar potential of such a model may have (before one uses the liberty to redefine the fields of
the model) 54 independent real parameters, as opposed to the two parameters of the SM scalar potential
or 14 free parameters of the general 2HDM scalar sector. The imposition of
global---discrete or continuous---symmetries on this model is therefore a good idea. The proposal by Weinberg  of a 3HDM
equipped with natural flavor conservation in the Yukawa sector \cite{Weinberg:1976hu}, for instance, included two discrete
symmetries generated by $\phi_2 \rightarrow -\phi_2$ and, separately, $\phi_3 \rightarrow -\phi_3$,
making the symmetry group of the model $\Z_2 \times \Z_2$.
This symmetry is then spontaneously broken by the vacuum expectation values (vevs) in the Higgs doublets. 
That model had a total of 18 real parameters.
Our approach is based not on family symmetries like the $Z_2$ ones described in the above example, but rather on
{\em generalized $CP$ symmetries} (or GCP), which relate the fields with their complex conjugates. For the reader's convenience and
also to set up the notation, we begin with a general reminder on the freedom in choosing $CP$ transformations
one has when building a model.

\subsection{The freedom of defining $CP$-symmetries}
\label{section:CP-freedom}

A self-consistent local quantum field theory does not uniquely specify how discrete symmetries,
such as $C$ and $P$, act on field operators \cite{feinberg-weinberg,lee-wick,book,weinberg-vol1}.
There is a great amount of freedom in defining these transformations, which becomes especially large
in the case of several fields with equal quantum numbers. This is due to the fact that such fields are
not physical by themselves; only the mass eigenstates obtained after spontaneous symmetry breaking
will correspond to physical particles. Any linear combination of those fields which preserves
the kinetic terms of the model will be equally acceptable as a {\em basis choice} for the theory.
Therefore, any symmetry of the Lagrangian which is supposed to incorporate a physically measurable
property, is defined up to an unconstrained basis choice shift.

Focusing now on a $CP$ transformation acting on several scalar fields $\phi_i$, $i = 1, \dots, N$,
one often considers the following GCP transformations \cite{gcp-standard,Grimus:1995zi}:
\be
J_X:\quad  \phi_i(\bx, t) \toCP {\cal CP}\,\phi_i(\bx, t)\, {\cal CP}^{-1} = X_{ij}\phi_j^*(-\bx, t), \quad X_{ij} \in U(N)\,.
\label{GCP}
\ee
If there exists a unitary matrix $X$ such that the Lagrangian of a model is invariant under this GCP
transformation, then the model is explicitly $CP$-conserving and $J_X$ can play the role of
``the $CP$-symmetry'' of the model \cite{book}. Notice that the ``conventional'' definition of $CP$
with $X_{ij} = \delta_{ij}$, so that $\phi_i(\bx, t) \toCP \phi_i^*(-\bx, t)$, is only one of many possible
definitions and is, in fact, a basis-dependent choice.

We find this terminological issue so important that, in abuse of the reader's patience, 
we spell it out once again. When we say that the model is $CP$-conserving,
we may refer to {\em any} form of GCP symmetry (\ref{GCP}), with whatever fancy $X$.
In particular, of the ``conventional'' definition of $CP$ transformations fails to leave
the lagrangian invariant, but a more complicated GCP transformation does, then the model is still $CP$-conserving
in the very traditional sense that all $CP$-odd observables are zero.
It is only when {\em none} of the transformations (\ref{GCP}) is a symmetry of the model
that we say that $CP$ violation takes place \cite{book}. 

The shape of the $X$ matrix
may have important consequences for the phenomenological behaviour of the models.
In the 2HDM, for instance, {\em three} different (and relevant)
choices for $X$ are possible, each leading to different accidental symmetries,
and three different $CP$-invariant models with vastly different phenomenologies
emerge from those choices \cite{Ivanov:2007de,2HDM-GCP}.

Notice now that applying $J_X$ twice generates a pure family transformation:
\be
\phi_i(\bx, t) \to ({\cal CP})^2\phi_i(\bx, t) ({\cal CP})^{-2} = (XX^*)_{ij}\phi_j(\bx, t)\,.
\label{GCP2}
\ee
Using the redefinition freedom one has in the choice of the basis of scalar fields, it is possible
to bring the matrix $X$ to a block-diagonal form \cite{weinberg-vol1,gcp-standard}, with the blocks
being either $1\times 1$ phases or $2\times 2$ matrices of the following type:
\be
\mmatrix{c_\alpha}{s_\alpha}{-s_\alpha}{c_\alpha}\quad \mbox{as in Ref.~\cite{gcp-standard},}\quad \mbox{or}\quad
\mmatrix{0}{e^{i\alpha}}{e^{-i\alpha}}{0}\quad \mbox{as in Ref.~\cite{weinberg-vol1}.}\label{block}
\ee
This is the simplest form of $X$ one can achieve with basis transformations in the scalar space $\mathbb{C}^N$.
If $X$ contains at least one $2\times 2$ block with $\alpha \not = 0$ or $\pi$, then $(J_X)^2 = XX^* \not = \mathbb{I}$.
This then means that the $CP$ transformation \eqref{GCP} is not an order-2 transformation. If $k$ is the smallest
integer such $(J_X)^k = \mathbb{I}$, the GCP transformation $J_X$ is said to be of order $k$.

One immediately sees that $k$ is necessarily an even number: one needs to perform conjugation
an even number of times to obtain the identity transformation. However, imposing the GCP of a {\em generic} even order $k$
immediately leads to accidental symmetries, including the GCP of a smaller order. Indeed, if $k$ has prime factors other than two,
one can factor them out and obtain a smaller-order GCP. The only way to prevent this possibility is to take $k = 2^p$, with $p\geq 1$
an integer number. Which means that the usual $CP$ is of order two (CP2), the first non-trivial higher-order $CP$ symmetry is CP4, the
next one is CP8, and so on.

Here we would like to reiterate again the point made three paragraphs earlier.
When we label a model as $CP$-conserving or $CP$-violating, we do not need to specify
whether it conserves or violates CP2, CP4, or a higher-order GCP.
$CP$-odd observables do not distinguish them.
If there exists at least one GCP transformation that leaves the model invariant,
then it is $CP$-conserving. Conversely, when we speak of $CP$-violation, we mean that all 
possible GCP transformations fail to leave the model invariant.

Although the $CP$-odd observables do not distinguish different classes of GCP transformations (CP2, CP4, etc),
the parameters of the lagrangian definitely do.
Since higher-order GCPs involve transformation between a pair of fields,
imposing it will certainly constrain the parameters stronger than the conventional $CP$ or, in general, any expression for CP2.
As a result, it may happen that the model itself does not offer enough freedom to implement a higher-order $CP$ symmetry.
For example, imposing a higher-order $CP$ symmetry on the scalar potential of the 2HDM produces accidental symmetries,
which include the usual $CP$ \cite{2HDM-GCP}. Thus, imposing higher-order $CP$ symmetry has always been viewed as
a compact way of defining a model, but not as a path towards {\em new} models that could not be achieved through the
usual ``order-2 $CP$ $+$ family symmetry'' combination. A rare exception is discussed in Ref.~\cite{trautner}, where
the higher-order $CP$ symmetries were classified as distinct opportunities for model building.
Further, extending these GCP symmetries to the Yukawa sector within 2HDM was
problematic \cite{Maniatis:2007de,Maniatis:2009vp,Ferreira:2010bm}, as they ran into trouble when confronted
with the experimental data (predicting some massless fermions and an insufficient CPV, for instance).
In a sense, 2HDM does not offer the model builder enough room to fully incorporate such a strongly constraining
symmetry as CP4, and one needs to extend the number of doublets to at least three.

\subsection{The scalar potential}

How many different global symmetries can one impose upon a given BSM model? Given the basis
redefinition freedom present in many such models, apparently different symmetries may, in fact,
be related by basis choices. For instance, in the 2HDM a symmetry of the form $\phi_1 \leftrightarrow
\phi_2$ is equivalent, in a different basis, to the usual $Z_2$ symmetry, $\phi_1 \rightarrow \phi_1$
and $\phi_2 \rightarrow -\phi_2$. In each basis, however, the Lagrangian of the model looks completely
different, with seemingly diverse relations between parameters. In the 2HDM, the work of
Ref.~\cite{Ivanov:2007de} proved that there are only six different symmetry classes, which
are not related among themselves by basis choices.

In the 3HDM the situation is more complicated.
In Ref.~\cite{Ferreira:2008zy} a first attempt at finding different classes of the 3HDM symmetries was
undertaken, but that study has been restricted to simple Abelian groups. A systematic and constructive
search for all discrete symmetry groups in the scalar sector of the 3HDM was performed in Ref.~\cite{abelian}
for Abelian and in Ref.~\cite{Ivanov:2012ry,Ivanov:2012fp} for discrete non-Abelian groups.
In each case, it was checked whether the family symmetry group can be further enlarged to include
a general $CP$ symmetry without producing any further accidental group.

This construction showed that, up to a basis choice, there exists only one 3HDM with a $CP$ symmetry of higher order,
to be specific, CP4, which does not lead to accidental symmetries. In the suitable basis, CP4 acts on the three Higgs
doublets in the following way:
\be
J: \quad \phi_i \toCP X_{ij} \phi_j^*\,,\quad
X =  \left(\begin{array}{ccc}
1 & 0 & 0 \\
0 & 0 & i  \\
0 & -i & 0
\end{array}\right)\,.
\label{J-def}
\ee
The most general potential respecting this CP4 symmetry can be written as $V=V_0+V_1$, where
\bea
V_0 &=& - m_{11}^2 (\phi_1^\dagger \phi_1) - m_{22}^2 (\phi_2^\dagger \phi_2 + \phi_3^\dagger \phi_3)
+ \lambda_1 (\phi_1^\dagger \phi_1)^2 + \lambda_2 \left[(\phi_2^\dagger \phi_2)^2 + (\phi_3^\dagger \phi_3)^2\right]
\nonumber\\
&+& \lambda_3 (\phi_1^\dagger \phi_1) (\phi_2^\dagger \phi_2 + \phi_3^\dagger \phi_3)
+ \lambda'_3 (\phi_2^\dagger \phi_2) (\phi_3^\dagger \phi_3)\nonumber\\
&+& \lambda_4 \left[(\phi_1^\dagger \phi_2)(\phi_2^\dagger \phi_1) + (\phi_1^\dagger \phi_3)(\phi_3^\dagger \phi_1)\right]
+ \lambda'_4 (\phi_2^\dagger \phi_3)(\phi_3^\dagger \phi_2)\,,
\label{V0}
\eea
with all parameters being necessarily real, and
\be
V_1 = \lambda_5 (\phi_3^\dagger\phi_1)(\phi_2^\dagger\phi_1) +
\lambda_8(\phi_2^\dagger \phi_3)^2 + \lambda_9(\phi_2^\dagger\phi_3)(\phi_2^\dagger\phi_2-\phi_3^\dagger\phi_3) + h.c.
\label{V1a}
\ee
with real $\lambda_5$ and complex $\lambda_8$, $\lambda_9$\footnote{In fact, in $V_1$ one can write additional terms
invariant under the same GCP transformation \eqref{J-def}, which was indeed done in the previous publications on this
model \cite{Ivanov:2015mwl,Aranda:2016qmp}. However, using the residual freedom of basis transformations which leave $J$
invariant, one can simplify $V_1$ to the form of Eq.~\eqref{V1a}. We provide an explanation of this procedure in
Appendix~\ref{appendix-simplifying}.}. 

Applying the transformation~\eqref{J-def} twice leads to $J^2 = X X^* = \mathrm{diag}(1,\,-1,\,-1)
\not = \mathbb{I}$. It is trivial to see that one recovers the identity transformation only after
applying $J$ {\em four} times: $J^4 = \mathbb{I}$. Thus, the transformation $J$ is indeed a GCP of order 4.
For generic values of the coefficients, this potential has no other Higgs-family or GCP
symmetries, apart from powers of $J$ \cite{abelian}.
In particular, this potential is not invariant under the ``conventional'' $CP$-symmetry or, in general,
under any other CP2. Nevertheless, the model is still $CP$-conserving because there exists at least one GCP
(namely, CP4) which is a symmetry of the model. The fact that the potential has no CP2 symmetry
is just irrelevant.

At this point, notice that there is no basis transformation that would make all
the coefficients of the scalar potential $V$ real \cite{Ivanov:2015mwl}.
Indeed, if it were possible to find such a real basis, then the potential would have an order-2 GCP. But such a symmetry
is absent in the CP4 3HDM; therefore, the real basis does not exist. The absence of the real basis does not contradict
explicit $CP$-conservation (and therefore, explicit $T$-conservation), because all basis-invariant combinations
of the scalar couplings are $CP$-even. This model completely settles the issue of whether explicit $CP$ conservation
is equivalent to the existence of a real basis \cite{Gunion-Haber}: they are equivalent only for $CP$ symmetries of order 2
and not for higher-order GCP.

As is conventional practice in building the models with extended Higgs sectors, it is necessary to require that
the quartic parameters $\lambda_i$ are such that the potential is bounded from below (BFB).
In other words, for quasiclassically large values of the Higgs fields along any direction in the scalar space,
the potential must rise to plus infinity. One usually assumes the strong version of the BFB condition, which requires
the quartic potential to strictly grow in any direction\footnote{In principle, potentials stable in a weak sense,
in which flat directions of their quartic potential are protected by the growing quadratic terms, are also acceptable.
However, they correspond to measure zero regions in the parameter space, and we can avoid them in
the phenomenological analysis.}. Necessary and sufficient conditions for BFB were obtained for the 2HDM earlier
in Ref.~\cite{Ivanov:2006yq}, but no such deduction has hitherto been possible for the general 3HDM, or even for the
CP4 3HDM we are dealing with. Nonetheless, we established in Appendix~\ref{appendix-BFB-conditions} a set
of {\em sufficient} BFB conditions, which, although somewhat overly restrictive, will guarantee that the potential
is indeed bounded from below. We will apply these conditions in our numerical analysis later on, which will
therefore be a conservative one.

\subsection{Extrema: generic solutions}

When a scalar potential is explicitly $CP$-conserving, the vacuum of the theory can preserve
$CP$, or spontaneously break it. The 2HDM, for instance, was first conceived as a model wherein
spontaneous CPV has occurred~\cite{Lee:1973iz}. The CP4 3HDM potential introduced
in the previous section is explicitly $CP$ conserving. Since we aim at extending the CP4 symmetry
to the Yukawa sector, it must be spontaneously broken; otherwise, the model would feature
mass-degenerate fermions \cite{Aranda:2016qmp}. Without lack of generality, one can write
the most generic charge-preserving vevs as
\be
\sqrt{2}\lr{\phi_i^0} = (v_1,\, v_2 e^{i\gamma_2},\, v_3 e^{i\gamma_3}) \equiv
(v_1,\, u c_\psi e^{i\gamma_2},\, u s_\psi e^{i\gamma_3})\,,
\label{vevs1}
\ee
where $v_1 > 0$, $u \equiv \sqrt{v_2^2 + v_3^2}$ and we used the standard notation $c_\psi \equiv \cos\psi$,
$s_\psi \equiv \sin\psi$. Later on, we will also use $t_\psi \equiv \tan\psi$.
We then expand the doublets around the extremum using the following conventions:
\be
\phi_1 = {1\over\sqrt{2}}\doublet{\sqrt{2}h_1^+}{v_1 + h_1 + i a_1},\
\phi_2 = {e^{i\gamma_2}\over\sqrt{2}}\doublet{\sqrt{2}h_2^+}{v_2 + h_2 + i a_2},\
\phi_3 = {e^{i\gamma_3}\over\sqrt{2}}\doublet{\sqrt{2}h_3^+}{v_3 + h_3 + i a_3}.\label{expansion}
\ee

By substituting this expansion in the Higgs potential $V$ and setting the coefficients of the linear terms
to zero we obtain the minimisation equations, also known as the tadpole conditions.
The coefficient of the linear term in $a_1$ gives us the following relation:
\be
{\lambda_5 \over 2} v_1 u^2 s_{2\psi} \sin(\gamma_2 + \gamma_3) = 0\,.\label{tadpole-a1}
\ee
Let us for the moment consider the generic situation with $\sin (2\psi) \not = 0$.
It leads to $\gamma_3 = - \gamma_2 \equiv - \gamma$.
The tadpole conditions for $a_2$ and $a_3$ then produce an additional relation,
\be
u^2\left(|\lambda_8| s_{2\psi}\sin\left[\arg(\lambda_8) - 4\gamma\right] + |\lambda_9|c_{2\psi}
\sin\left[\arg(\lambda_9) - 2\gamma \right]\right)  = 0\,.
\label{tadpole-a2a3}
\ee
For given $\lambda_8$ and $\lambda_9$, this equation relates the phase $\gamma$ with the angle $\psi$.
In order to simplify the analysis, we find it convenient to switch now to the {\em real vev basis}
by rephasing $\phi_2 \to \phi_2 e^{-i\gamma}$ and $\phi_3 \to \phi_3 e^{i\gamma}$.
With this basis transformation, all real parameters in the potential stay the same, while $\lambda_8$
and $\lambda_9$ are rephased so that the quantity $2\arg (\lambda_9) - \arg(\lambda_8)$ remains unchanged.
In the real vev basis, the tadpole condition above is written as
\be
s_{2\psi} \Im (\lambda_8) + c_{2\psi}\Im (\lambda_9) = 0\quad \Leftrightarrow\quad
\tan 2\psi = {2v_2v_3 \over v_2^2 - v_3^2} = - {\Im (\lambda_9) \over \Im (\lambda_8)}\,.
\label{t2b}
\ee
Note that the latter can not be considered as an expression for $\psi$ in terms of the parameters of
the original potential since the phases of $\lambda_8$ and $\lambda_9$ depend now on $\gamma$.
However, if $\Im(\lambda_8)$ is a free parameter and if $\psi$ is known, one can deduce $\Im(\lambda_9)$.

The tadpoles for $h_i$ lead to the following three relations:
\bea
m_{11}^2 &=& \lambda_1 v_1^2 + {1 \over 2} u^2\lambda_{34} + {1 \over 2} u^2\lambda_5\, s_{2\psi}\,,
\label{m11}\\
m_{22}^2 &=& \lambda_2 u^2 + {1 \over 2}v_1^2 \lambda_{34}
+ {1\over 2}  u^2 \Re(\lambda_9)\, t_{2\psi}\,,
\label{m22}\\
0 &=&\lambda_5 v_1^2 c_{2\psi} + \Lambda u^2 s_{2\psi} c_{2\psi} + \Re (\lambda_9)\, u^2 c_{4\psi}\,,
\label{t23}
\eea
where we used the following shorthand notations:
\be
\Lambda \equiv {\lambda'_{34}\over 2} + \Re (\lambda_8) - \lambda_2\,,
\quad \lambda_{34} \equiv \lambda_3 + \lambda_4\,, \quad \lambda'_{34} \equiv \lambda'_3 + \lambda'_4\,.
\label{Lambda}
\ee
In total, therefore, we have five minimisation conditions to solve. One of them produces a relationship between
the phases of the vevs of $\phi_2$ and $\phi_3$, the remaining four must still be solved. As is usually the case
in multi-Higgs models, obtaining the analytical expressions for the vevs (and their phases) in terms of the potential's
parameters is exceedingly difficult. It is much simpler, in a numerical study, to take the values of the vevs and
their phases as inputs. Therefore, we consider $\psi$ as a {\em free parameter} and use Eq.~\eqref{t23} and
$\sqrt{v_1^2 + u^2} \equiv v =$ 246 GeV to determine $v_1^2$ and $u^2$ in terms of $\psi$ and the scalar
quartic couplings. Then, Eqs.~\eqref{m11} and \eqref{m22} may be used to extract $m_{11}^2$ and $m_{22}^2$.
We will follow a similar procedure bearing also in mind our desire to find an acceptable scalar (and fermionic)
mass spectrum.

If $\psi$ is considered as input in a numerical scan over parameter space of the model, one needs to specify
its range. Here, we argue that $0 \le \psi \le \pi/2$, which corresponds to positive $v_2$ and $v_3$, faithfully covers
the entire set of relevant cases. The arguments go as follows. When the discrete symmetry CP4 is spontaneously broken,
the potential has four degenerate minima, all related by the broken symmetry transformations but all corresponding
to the same physics. These four minima are obtained by consecutively applying the transformation
$v_2 \leftrightarrow v_3$, $\gamma \to \gamma + \pi/2$, while the real $v_1$ stays unchanged.
Now, suppose we allow the free parameter $\psi$ to take any value, thus allowing $v_2$ and $v_3$ to be
either positive or negative. By the previous argument, we immediately see that the point $(v_1, -|v_2|, -|v_3|)$
is two transformations away from $(v_1, |v_2|, |v_3|)$, leading to the same model. The point $(v_1, |v_2|, -|v_3|)$
represents a real-vev basis transformation of the point $(v_1, i|v_2|, -i|v_3|)$, which, in turn, corresponds to
the same model as $(v_1, |v_3|, |v_2|)$. Therefore, whatever value $\psi$ takes, the model it leads to can also
be found in the first quadrant of $\psi$.

\subsection{Extrema: special points}

Let us now consider two special values of $\psi$. The first case is when $s_{2\psi} = 0$, and without loss of generality,
we can set $\psi = 0$ meaning $v_3 = 0$. In this case, the tadpole condition \eqref{tadpole-a1} is also satisfied.
The other tadpole conditions get simplified, and after some algebra we arrive at the following relations:
\be
\Im(\lambda_9) = 0\,,\quad |\lambda_5| v_1^2 = |\lambda_9| u^2\,,\quad
m_{11}^2 = \lambda_1 v_1^2 + {1\over 2}\lambda_{34}u^2\,,\quad
m_{22}^2 = \lambda_2 u^2 + {1\over 2}\lambda_{34}v_1^2\,. \label{s2b-zero}
\ee
These relations are exactly those which we would get from the previous subsection in the limit $s_{2\psi} \to 0$.
Thus, we do not need to include this point as a separate case; we simply allow $\psi$ to start from zero.

The second singular point is $c_{2\psi} = 0$, implying $v_2 = v_3$. Then, the tadpole condition \eqref{tadpole-a2a3}
can be satisfied, in the real vev basis, either when $u = 0$ or when $\lambda_8$ is real, while $\lambda_9$ is purely
imaginary. The former option would lead to an unphysical fermion spectrum, as we have already mentioned,
while in the latter case the symmetry content of the model increases and involves now several accidental symmetries
including the $CP$ symmetry of order 2. This model, not being CP4-driven, falls beyond the scope of the present study.
However the points in the vicinity of this limit, $c_{2\psi} \ll 1$, are acceptable and, when accompanied with
correspondingly small $\Re(\lambda_9)$, can potentially lead to realistic models.

\subsection{Scalar mass matrices}

In the real vev basis, we can expand the potential around the chosen extremum up to quadratic
terms and construct the charged and neutral scalar mass matrices. For charged scalar fields, we get
terms of the form $h_i^- ({\cal M}_{ch})_{ij} h_j^+$, where
\be
{\cal M}_{ch} = {1\over 2}\mmmatrix{-u^2(\lambda_4 + \lambda_5 s_{2\psi})}%
{v_1u(\lambda_4 c_\psi + \lambda_5 s_\psi)}{v_1u(\lambda_4 s_\psi + \lambda_5 c_\psi)}%
{\cdot }{u^2 s_\psi^2 \tilde\Lambda -\lambda_4 v_1^2}%
{- u^2 s_\psi c_\psi \tilde\Lambda - \lambda_5 v_1^2}%
{\cdot}{\cdot}%
{u^2 c_\psi^2 \tilde\Lambda - \lambda_4 v_1^2}\,,\label{mass-matrix-charged}
\ee
with $\tilde\Lambda \equiv \lambda'_3 - 2\lambda_2 - 2 \Re(\lambda_9)\, t_{2\psi}$.
Here, for simplicity, the dots indicate the duplicated entries of this symmetric matrix.
One can explicitly verify that the charged Goldstone boson, which corresponds to the
combination of fields given by
\be
G^\pm = (v_1 h_1^\pm + u c_\psi h_2^\pm + u s_\psi h_3^\pm)/v
\ee
is an eigenvector of this matrix with zero eigenvalue, as expected.
The masses of the physical charged Higgs bosons can be explicitly calculated
from the traces of powers of this matrix as usual,
\bea
m^2_{H_1^\pm} + m^2_{H_2^\pm} \equiv \Tr {\cal M}_{ch}
&=& -\lambda_4 v_1^2 - {1\over 2}u^2(\lambda_4 + \lambda_5 s_{2\psi} - \tilde\Lambda)\,,
\label{m2ch-1} \\
2m^2_{H_1^\pm}m^2_{H_2^\pm} \equiv (\Tr {\cal M}_{ch})^2 -\Tr({\cal M}_{ch}^2) &=&
{1\over 2}v^2 \left[v_1^2 (\lambda_4^2 - \lambda_5^2) -
u^2 \tilde\Lambda (\lambda_4 + \lambda_5 s_{2\psi})\right]\,.
\label{m2ch-2}
\eea
For the neutral scalars, the fact that $CP$ is spontaneously broken by the vacuum implies the presence
of a mixing between the real and imaginary neutral components of the Higgs doublets, $h_i$ and $a_j$. Thus,
the resulting $6\times 6$ mass matrix can be written via symmetric 3-by-3 blocks,
\be
{\cal M}_n = \mmatrix{M_{h}}{M_{ha}}{M_{ha}^T}{M_{a}}\,,
\label{neutralMM}
\ee
where
\bea
M_h &=&
\mmmatrix{2\lambda_1 v_1^2}{\lambda_{34}v_1 u \, c_\psi }{\lambda_{34}v_1 u \,s_\psi}
{\cdot}%
{2\lambda_2u^2\, c_\psi^2}%
{2\lambda_2 u^2\, s_\psi c_\psi}
{\cdot}%
{\cdot}%
{2\lambda_2 u^2\, s_\psi^2} +
\mmmatrix{0}{\lambda_5 v_1 u\, s_\psi}{\lambda_5 v_1 u\, c_\psi}
{\cdot}%
{\Lambda u^2\, s_\psi^2}%
{-\Lambda u^2\, s_ \psi c_\psi - \lambda_5 v_1^2}
{\cdot}%
{\cdot}%
{\Lambda u^2\, c_\psi^2} \nonumber\\
&&+ {1\over 2}\Re \lambda_9\, u^2\, t_{2\psi}
\mmmatrix{0}{0}{0}{\cdot}{3c_{2\psi}-1}{3s_{2\psi}}{\cdot}{\cdot}{-3c_{2\psi}-1}\,,\label{mhh}\\
M_a &=& \mmmatrix{-\lambda_5 u^2 s_{2\psi}}{\lambda_5 v_1 u s_\psi}{\lambda_5 v_1 u c_\psi}%
{\cdot}{u^2 \Lambda' s_\psi^2}{- \lambda_5 v_1^2 - \Lambda' u^2 s_\psi c_\psi}%
{\cdot}{\cdot}{u^2 \Lambda' c_\psi^2}\,,\label{maa}
\eea
with $\Lambda' \equiv \Lambda - 2\Re (\lambda_8) - \Re(\lambda_9)\, t_{2\psi}$.
Finally, the $h/a$ mixing terms depend on the imaginary parts of $\lambda_8$ and $\lambda_9$,
as could be expected. Using Eq.~\eqref{t2b}, we can parametrize these coefficients as
\be
\Im(\lambda_8) = c_{2\psi} \lambda_{89}\,, \quad \Im(\lambda_9) = -s_{2\psi} \lambda_{89}\,,
\quad
\lambda_{89} \equiv \sqrt{(\Im \lambda_8)^2 + (\Im\lambda_9)^2}\,.
\label{lam89}
\ee
Then, the mixing terms can be grouped as $- \lambda_{89} u^2 (s_\psi h_2 - c_\psi h_3)(s_\psi a_2 - c_\psi a_3)$,
which implies that the off-diagonal block in Eq.~\eqref{neutralMM} is given by
\be
M_{ha} = - \lambda_{89} u^2 \mmmatrix{0}{0}{0}{0}{s_\psi^2}{-s_\psi c_\psi}{0}{-s_\psi c_\psi}{c_\psi^2}\,.
\label{mha}
\ee
One can again explicitly check that the neutral Goldstone boson, given by the following combination of
fields
\be
G^0 = (v_1 a_1 + u c_\psi a_2 + u s_\psi a_3)/v\,,
\label{neutral-goldstone}
\ee
is an eigenvector of the neutral mass matrix ${\cal M}$ with a zero eigenvalue.
The resulting five neutral physical Higgs bosons mix and do not possess definite $CP$-properties, as
should be expected considering that the $CP$ symmetry has been broken.
In particular, none of the neutral scalar states possess the parities or $CP$-charges defined in
Refs.~\cite{Ivanov:2015mwl,Aranda:2016qmp} for the CP4 unbroken case.

As shown in Refs.~\cite{Ivanov:2015mwl,Aranda:2016qmp}, in the case of a vacuum with unbroken CP4,
{\em i.e.}~with $u=0$, the physical Higgs spectrum organizes itself in mass-degenerate pairs, thus resembling
that of the 2HDM. In such a vacuum, the model has a pair of mass-degenerate charged scalars,
one SM-like neutral Higgs and two pairs of mass-degenerate neutral scalars, with masses $m$ and $M$
(one is an analogue of the heavier $CP$-even scalar in the 2HDM, $H$, the other is an analogue of the 2HDM
pseudoscalar state, $A$). The spontaneous breaking of CP4 will induce a splitting in the mass spectrum,
which vanishes in the $u \to 0$ limit, assuming that the quartic couplings remain fixed.

\subsection{Scalar alignment limit}
\label{section-scalar-alignment}

An important property of viable multi-Higgs models in the post LHC era is that they should have regions in
their parameter space for which one of their neutral scalars possesses a mass of about 125 GeV and closely
resembles the SM Higgs boson. By this statement we mean that the couplings of this scalar
to the SM gauge bosons and fermions must be very similar in magnitude to the corresponding
SM values for those couplings. This may be achieved either in a ``natural'' way via symmetries (like
in the case of the inert 2HDM \cite{Deshpande:1977rw,Barbieri:2006dq}), or by
a fine-tuning of the multi-Higgs model considered. In many scenarios, the desired region of
parameter space can arise more or less ``naturally'' in a {\em decoupling} regime, wherein
the extra scalar states are much heavier than the lightest SM-like one 
(see Ref.~\cite{Gunion:2002zf} for its introduction to the 2HDM).
An intermediate case is the {\em alignment limit}, where some extra scalar masses may be low 
if certain relations by the couplings are satisfied. 
Again, within the 2HDM, the alignment without decoupling, and possible symmetry-based pathways to it,
were considered in \cite{Dev:2014yca,Bernon:2015qea,Bernon:2015wef,Grzadkowski:2016szj}
and for maximally symmetric models beyond two doublets \cite{Pilaftsis:2016erj}.

Let us examine how {\em exact scalar alignment} may arise in our model.
In the original basis, the vevs of the doublets are given by Eq.~\eqref{vevs1}, which we now rewrite as follows
\be
\sqrt{2}\lr{\phi_i^0} = (v_1,\, v_2 e^{i\gamma},\, v_3 e^{-i\gamma})
\equiv v\, (c_\beta, s_\beta c_\psi e^{i\gamma}, s_\beta s_\psi e^{-i\gamma})\,.\label{vevs2}
\ee
The {\em Higgs basis} is defined as a basis in which only the first doublet gets a vev. In that basis,
we write the neutral complex fields (lower components of the Higgs doublets) as
\be
\Phi_i = \triplet{\Phi_1}{\Phi_2}{\Phi_3} \equiv {1\over\sqrt{2}}\triplet{\rho_1 +
i G^0}{\rho_2 + i \eta_2}{\rho_3 + i \eta_3}\,,
\quad \lr{\Phi_i} = {1\over\sqrt{2}}\triplet{v}{0}{0}\,.\label{Higgs-basis}
\ee
In general, the fields $\rho_i$ and $\eta_i$ are not mass eigenstates, and the neutral mass matrix
includes mixing terms between them. The exact {\em scalar alignment} refers to the situation where one of the states,
e.g. $\rho_1$, is a mass eigenstate, identified with the 125 GeV Higgs boson $H_{125}$. If this is the case, then its tree-level
couplings to the $W$ and $Z$ bosons and to fermions are exactly the same as in the SM. 
Therefore, in such alignment limit the SM-like state will not mediate any tree-level FCNCs. 
The other (heavier) neutral scalars can, and generally do, have tree-level FCNCs, as will be
discussed in Section~\ref{section:yukawa-sector}.

To find the condition for scalar alignment, it is necessary to study the neutral scalar mass matrix
in the Higgs basis. To do so, we first remark that the Higgs basis is not uniquely defined.
In the 3HDM, any unitary transformation between $\Phi_2$ and $\Phi_3$ in Eq.~\eqref{Higgs-basis} preserves
the definition of the Higgs basis\footnote{A similar $\Phi_2$ rephasing freedom has already been noticed in the 2HDM.}.
We make the following traditional (and convenient) choice for the Higgs basis:
\be
\triplet{\Phi_1}{\Phi_2}{\Phi_3} =
\mmmatrix{c_\beta}{s_\beta c_\psi}{s_\beta s_\psi}{0}{-s_\psi}{c_\psi}{s_\beta}{-c_\beta c_\psi}{-c_\beta s_\psi}
\triplet{\phi_1}{\phi_2 e^{-i\gamma}}{\phi_3 e^{i\gamma}}\,.\label{matrix-P}
\ee
If one starts in the real vev basis, one should just set the phase $\gamma$ to zero.
For completeness, we give in Appendix~\ref{appendix-three-bases} all expressions in the original basis
with a non-zero $\gamma$.

To proceed to the physical scalars, we switch to the 6-dimensional real scalar field space
and consider the $6\times 6$ neutral mass matrix ${\cal M}$, see Eq.~\eqref{neutralMM}.
We order the fields in the real vev basis as $\varphi_a = (h_1, h_2, h_3, a_1, a_2, a_3)$
and in the chosen Higgs basis as $\Phi_a = (G^0, \rho_1, \rho_2, \rho_3, \eta_2, \eta_3)$.
The rotation to the Higgs basis is done by $\Phi_a = P_{ab}\varphi_b$, where the $6\times 6$
matrix $P_{ab}$ is given by Eq.~\eqref{rotation-matrix-P}. To obtain the form of the neutral mass
matrix in the Higgs basis, we start from ${\cal M}$ from Eq.~\eqref{neutralMM} and use the rotation
matrix $P$ so that ${\cal M}^H = P {\cal M} P^T$. With this rotation, one immediately finds that
the first column and the first row of ${\cal M}^H$ only contain zeros, by virtue of the neutral
Goldstone decoupling.

The exact Higgs alignment happens when the {\em second} row and column also decouple
from the rest, i.e.
\be
{\cal M}^H = \mmmatrix{0}{0}{0_4}{0}{m^2_{H_{125}}}{0_4}{0_4}{0_4}{{\cal M}^H_{4\times 4}}\,.
\ee
Therefore, in order to establish the scalar alignment conditions, one needs to calculate the second row
of ${\cal M}^H$ and to set its off-diagonal elements to zero. We did that and observed that
${\cal M}^H_{23} = {\cal M}^H_{25} = {\cal M}^H_{26} = 0$ are automatically fulfilled for
generic vevs $v_1$, $u$ and angle $\psi$. The only non-zero entries of the second row can be
written, after some algebra, in a very compact form:
\be
{\cal M}^H_{22} = 2c_\beta^2 m_{11}^2 + 2s_\beta^2 m_{22}^2 \,,\quad
{\cal M}^H_{24} = \sin{2\beta}\, (m_{11}^2 - m_{22}^2)\,.
\label{alignment-condition-0}
\ee
Exact scalar alignment is achieved when ${\cal M}^H_{24} = 0$, which implies
\be
m_{11}^2 = m_{22}^2\,,
\label{alignment-condition}
\ee
irrespectively of all other parameters of the model. Then the SM-like Higgs mass
becomes simply
\be
m_{H_{125}}^2 = 2 m_{11}^2\,.
\label{alignment-condition-SM-mass}
\ee
In this limit, no tree-level FCNCs occur in the $H_{125}$ interactions with fermions, and the couplings of this
scalar state to the gauge bosons are identical to those of the SM Higgs boson. This is an alignment without
decoupling: the remaining scalars can have any values of masses.

The simplicity of the scalar alignment condition \eqref{alignment-condition} is not surprising. 
Indeed, take {\em any} multi-Higgs-doublet potential with universal quadratic term
and rewrite it in terms of real fields $\varphi_a$,
\be
V = - m^2 \sum_a \varphi_a^2 + \Lambda_{abcd}\varphi_a \varphi_b\varphi_c\varphi_d \,,
\ee
with arbitrary $\Lambda_{abcd}$ constrained only by BFB conditions. 
Then this potential automatically incorporates the exact scalar alignment,
which can be verified by direct differentiation.
One scalar mass eigenstate is always aligned with the direction of vevs,
and its mass squared is $2m^2$. The non-trivial result of the above exercise is that, 
within CP4 3HDM with fully spontaneously broken CP4 symmetry,
setting $m_{11}^2 = m_{22}^2$ is the only way to impose scalar alignment. 


\section{CP4 symmetric Yukawa sector}
\label{section:yukawa-sector}

\subsection{Yukawa models with CP4}

The CP4 symmetry can be extended to the Yukawa sector, provided
the CP4 transformation also mixes the fermion generations, as follows
\be
\psi_i \toCP Y_{ij} \psi_j^{CP}\,, \quad\mbox{where} \quad \psi^{CP} = \gamma^0 C \bar\psi^T\,.
\label{fermion-GCP}
\ee
Such an extension has been performed in the framework of 2HDM in Refs.~\cite{Maniatis:2007de,
Maniatis:2009vp,Ferreira:2010bm} and ran into difficulties with excessive accidental symmetries.
In this work, we implement such an extension for the CP4 3HDM.

The CP4 symmetry strongly constrains the Yukawa interaction matrices. In Ref.~\cite{Aranda:2016qmp},
some examples of such interactions were given under the simplifying assumption that the right-handed up
and down fermions, as well as the left-handed doublets, transform in the same way, i.e. $Y^{L} = Y^{d} = Y^{u}$.
In this work, we lift this assumption and derive all possible forms of the Yukawa interaction matrices compatible
with CP4 and not running into immediate conflict with experiment, that is, not leading to massless fermions or
an insufficient mixing.

In this work, we focus on the quark sector only while the lepton sector can be incorporated in a similar way.
The quark Yukawa Lagrangian
\be
-{\cal L}_Y = \bar q_L \Gamma_a d_R \phi_a + \bar q_L \Delta_a u_R \phi_a^* + h.c.,
\ee
in which we explicitly indicated the Higgs family index and omitted for clarity the quark flavor indices,
is invariant under CP4 if and only if
\be
(Y^L)^\dagger \Gamma_a Y^d X_{ab} = \Gamma^*_b\,,\quad
(Y^L)^\dagger \Delta_a Y^u X^*_{ab} = \Delta^*_b\,.
\label{Yukawa-conditions}
\ee
With the explicit expression for $X$ given by Eq.~\eqref{J-def}, we get
\bea
&&
(Y^L)^\dagger \Gamma_1 Y^d = \Gamma_1^*\,,\quad
i (Y^L)^\dagger \Gamma_2 Y^d = \Gamma_3^*\,,\quad
-i (Y^L)^\dagger \Gamma_3 Y^d = \Gamma_2^*\,,\nonumber\\[1mm]
&&
(Y^L)^\dagger \Delta_1 Y^u = \Delta_1^*\,,\quad
-i (Y^L)^\dagger \Delta_2 Y^u = \Delta_3^*\,,\quad
i (Y^L)^\dagger \Delta_3 Y^u = \Delta_2^*\,.\label{Yukawa-conditions-2}
\eea
As usual, an appropriate change of the basis in the $q_L$, $u_R$, $d_R$ spaces
can bring all the matrices $Y$ to the block-diagonal form
\be
Y = \mmmatrix{0}{e^{i\alpha}}{0}{e^{-i\alpha}}{0}{0}{0}{0}{1}\,.\label{matrix-Y}
\ee
and we allow the parameters $\alpha_L$, $\alpha_d$, and $\alpha_u$ to be all different.
Here, we selected the third fermion to be a CP4 singlet but any other choice would be equivalent.

As we have seen above, in order to properly define inequivalent generalized $CP$ transformations 
on the fermion sector, the possible values of the phases in Eq.~\eqref{matrix-Y} must be such that 
the matrices $Y^{*}Y$ are of finite order $2^{p-1}$ with a distinct $p$ defining an inequivalent 
CP$2^{p}$ symmetry. When solving Eqns.~(\ref{Yukawa-conditions-2}), we do not assume that $\alpha$ corresponds 
to the CP4 case. In fact, those equations only imply that the fermion {\em bilinears} coupled to $\phi_2$ and $\phi_3$
must faithfully transform under CP4, but the transformation law of the fermion fields individually
may be of even higher order. We leave open this possibility and just solve the coupled equations.

We found that there are only four classes of Yukawa matrices $\Gamma_a$ and $\Delta_a$ satisfying Eq.~\eqref{Yukawa-conditions}
for some $\alpha$'s and not running into an immediate conflict with the data. We label them as cases $A, B_1, B_2, B_3$.
We describe them below for $\Gamma$'s in terms of their independent complex parameters $g_{ij}$, and then briefly
comment on matrices $\Delta_a$. Note that although for notational simplicity we use the same names for the parameters
$g_{ij}$ for the scenarios $A, B_1, B_2, B_3$, they should be regarded as different ones.

Before listing these results, one comment is in order. When solving all equations for $\alpha$'s and $\Gamma$'s, we often
obtain seemingly different solutions with extra minus factors in some rows or columns. We checked that in all cases they
represent the same model and differ just by a sign flip in one or several fermion fields. The four cases shown below are the ones
which cannot be reduced to one another.

\begin{itemize}
\item
{\bf Case $A$.} $e^{i\alpha_L} = 1$ and $e^{i\alpha_d} = 1$, giving
\be
\Gamma_1 = \mmmatrix{g_{11}}{g_{12}}{g_{13}}%
{g_{12}^*}{g_{11}^*}{g_{13}^*}%
{g_{31}}{g_{31}^*}{g_{33}}\,,\quad
\Gamma_{2,3} = 0\,,\label{caseA}
\ee
It may seem that $\Gamma_1$ is not as generic as it would be in the $CP$-conserving version of the SM
because the off-diagonal elements
are related to each other. However, when $\alpha = 0$, the CP-symmetry within the Yukawa sector
is effectively of order 2. As a consequence, there exists a fermion basis in which the CP-transformation
is the canonical one in the fermion sector and, in this basis, $\Gamma_1$ is simply an arbitrary real matrix.
\item
{\bf Case $B_1$.} $e^{i\alpha_L} = i$ and $e^{i\alpha_d} = 1$, giving
\be
\Gamma_1 = \mmmatrix{0}{0}{0}{0}{0}{0}{g_{31}}{g_{31}^*}{g_{33}}\,,\quad
\Gamma_2 = \mmmatrix{g_{11}}{g_{12}}{g_{13}}{g_{21}}{g_{22}}{g_{23}}{0}{0}{0}\,,\quad
\Gamma_3 =  \mmmatrix{-g_{22}^*}{-g_{21}^*}{-g_{23}^*}{g_{12}^*}{g_{11}^*}{g_{13}^*}{0}{0}{0}\,.
\label{caseB1}
\ee
\item
{\bf Case $B_2$.} $e^{i\alpha_L} = 1$ and $e^{i\alpha_d} = i$, giving
\be
\Gamma_1 = \mmmatrix{0}{0}{g_{13}}{0}{0}{g_{13}^*}{0}{0}{g_{33}}\,,\quad
\Gamma_2 = \mmmatrix{g_{11}}{g_{12}}{0}{g_{21}}{g_{22}}{0}{g_{31}}{g_{32}}{0}\,,\quad
\Gamma_3 =  \mmmatrix{g_{22}^*}{-g_{21}^*}{0}{g_{12}^*}{-g_{11}^*}{0}{g_{32}^*}{-g_{31}^*}{0}\,.
\label{caseB2}
\ee
\item
{\bf Case $B_3$.} $e^{i\alpha_L} = i$ and $e^{i\alpha_d} = i$, giving
\be
\Gamma_1 = \mmmatrix{g_{11}}{g_{12}}{0}{-g_{12}^* }{g_{11}^*}{0}{0}{0}{g_{33}}\,,\quad
\Gamma_2 = \mmmatrix{0}{0}{g_{13}}{0}{0}{g_{23}}{g_{31}}{g_{32}}{0}\,,\quad
\Gamma_3 = \mmmatrix{0}{0}{-g_{23}^*}{0}{0}{g_{13}^*}{g_{32}^*}{-g_{31}^*}{0}\,.
\label{caseB3}
\ee
\end{itemize}
All parameters apart from $g_{33}$ can be complex in all cases. Notice also that in all cases
the matrices $\Gamma_{2,3}$ are expressed in terms of the same complex parameters and
have the same textures.

For the up-quark sector, we get the same structures for $\Delta$'s. Indeed, the equations for $\Delta$'s are
the same as for $\Gamma$'s with the exchange $2\leftrightarrow 3$. However, when constructing a viable model,
we are not forced to select the same case for $\Gamma$'s and $\Delta$'s. We only must ensure that the transformation
properties of the left-handed doublets (i.e.~the values of $\alpha_L$) are the same in both sectors.
Therefore, we have two series of four
different combinations each for the down and up quark sectors:
\bea
\alpha_L = 0: && (A, A),\ (A, B_2),\ (B_2,A),\ (B_2, B_2),\\
\alpha_L = \pi/2: && (B_1, B_1),\ (B_1, B_3),\ (B_3, B_1),\ (B_3,B_3).
\label{combininig-cases}
\eea

Note, in the present analysis, we do not consider the case (A,A) which corresponds to a situation when the CP4 symmetry 
does not affect the fermion sector. The CKM matrix in this case is real at the tree level. However, the original $CP$ symmetry 
is broken by the Higgs sector giving a plausible chance to generate CPV effects in the fermion sector by means 
of radiative corrections. Whether this mechanism is capable of producing the experimentally observed amount of CPV
deserves a closer study that is left for future work.

By inspecting the textures of the Yukawa matrices in cases $B_1, B_2, B_3$, one immediately sees
that CP4 must be spontaneously broken. The unbroken case corresponds to the vev alignment $(v,0,0)$,
and in this case $M_d = \Gamma_1 v/\sqrt{2}$ will produce pathological quark masses in all cases B, by forcing some quarks
to be either massless or mass-degenerate \cite{Aranda:2016qmp}. Also, the strength of CP4 breaking cannot be too weak,
because it is the value of $u$ multiplied by some elements of $\Gamma_{2,3}$ that drives the mass splitting across
the two fermion generations. Fortunately, as was mentioned in the previous section, there is no reason to expect $u \ll v_1$ 
in a generic situation.

Let us also remark that when solving matrix equations \eqref{Yukawa-conditions-2} we found other non-trivial solutions.
We do not list them because they immediately enter in conflict with the quark masses/mixing properties.
For example, there exists a solution with $\alpha_L = \alpha_d = \alpha_u = \pi/4$ with the following Yukawa textures
\be
\Gamma_1, \Delta_1 = \mmmatrix{\times}{0}{0}{0}{\times}{0}{0}{0}{\times}\,,\quad
\Gamma_{2,3}, \Delta_{2,3} = \mmmatrix{0}{\times}{0}{\times}{0}{0}{0}{0}{0}\,.\label{YukawacaseD2}
\ee
As mentioned above, the fact that the phases in Eq.~\eqref{matrix-Y} are $\pi/4$ implies 
that the $CP$ symmetry acts on fermionic fields as CP8, and it only becomes CP4 for the fermion bilinears.
However, when multiplied by vevs, these Yukawa matrices lead to block-diagonal $M_d$ and $M_u$, both of which are diagonalized
by rotations between the first and second fermion generations. This means that $V_{CKM}$ will have a block-diagonal form
with only one mixing angle, which contradicts experimental results. Therefore, this solution and other similar cases are disregarded.

\subsection{Reconstructing Yukawa matrices}
\label{section-reconstructing-yukawas}

Having picked up one of the above cases and having fixed the vevs $v_i$, one can obtain the quark mass matrix. For example,
for the down quarks we have
\be
M_d = {1\over \sqrt{2}}\sum_{i=1}^3 v_i \Gamma_i\,.
\label{Md-via-Gammas}
\ee
Even if the $\Gamma$'s have a given structure, they all collapse to a single matrix $M_d$ where this structure may not 
be present anymore. Of course, in texture-constrained situations 
in which the Yukawa matrices of different Higgs fields do not share common non-zero entries, 
one can reconstruct $\Gamma_a$ from $M_d$. 
But in a generic situation with overlapping Yukawa matrices, this is impossible even knowing $v_i$. 

The remarkable property of the CP4 3HDM is that, despite overlapping entries, 
{\em it is possible} to reconstruct all $\Gamma$'s from $M_d$ once we know the vevs 
and which of the above cases $A, B_1, B_2, B_3$ we deal with. This is so because in all the cases 
there is a pair of entries of $M_d$ which is determined by two elements $g_{ij}$ only. Consider, for example, the case 
$B_3$ and the off-diagonal elements in the third column:
\be
\doublet{(M_d)_{13}}{(M_d)^*_{23}} = {1 \over\sqrt{2}}\mmatrix{v_2}{-v_3}{v_3}{v_2}\doublet{g_{13}}{g_{23}^*}\,.
\ee
By inverting this matrix relation, one reconstructs the elements $g_{13}$ and $g_{23}$ in terms of $(M_d)_{13}$ and $(M_d)_{23}$.
The same holds for other elements.

The resulting mass matrices for the models $B_1$, $B_2$ and $B_3$ are constrained only by the following relations:
$(M_{d})_{32} = (M_{d})^{*}_{31}$ for case $B_1$, $(M_{d})_{23} = (M_{d})^{*}_{13}$ for case $B_2$,
$(M_{d})_{22} = (M_{d})^{*}_{11}$ and $(M_{d})_{21} = -(M_{d})^{*}_{12}$ for case $B_3$.
However, in any of these cases, the matrices
\be
H_d \equiv M_d M_d^\dagger\,, \quad H_u = M_u M_u^\dagger\,,\label{HdHu}
\ee
are unconstrained hermitean matrices.
In the absence of any further restrictions over the domains of the independent 
matrix elements in $M_{d}$ and $M_{u}$, defined by the values of $ \Gamma_{a} $, $ \Delta_{a} $ and $ v_{i} $ in Eq. \eqref{Md-via-Gammas}, 
it is straightforward to reproduce all quark masses, mixing angles, and the $CP$ violation parameter. 
In many multi-Higgs-doublet models, the usual procedure $(\Gamma_a, \Delta_a) \to (M_d, M_u) \to (H_d, H_u) \to \{m_i, V_{CKM}\}$,
involves, through the first two steps, a map which does not cover the entire space of all hermitean $H_d$ and $H_u$.
As a result, one may fail to reproduce all quark masses, mixing angles, and the $CP$ violation parameter.
In CP4 3HDM, we see two remarkable properties:
\begin{itemize}
\item
all hermitean $H_d$ and $H_u$ are reachable for an appropriate choice of $(\Gamma_a, \Delta_a)$ and vevs;
\item
the first step $(\Gamma_a, \Delta_a) \to (M_d, M_u)$ is {\em invertible} once vevs are knows. 
\end{itemize}
If one wishes, one can perform the scan of the Yukawa parameters space starting with 
the physical quark masses, mixing, and CPV and then generating $g_{ij}$ elements which, by construction, will exactly reproduce 
the measured values. These features serve as an ``existence proof'' of good parameter space points,
in the sense that we will be able to fit, within CP4 3HDM, all the fermion masses, mixing,
and the amount of $CP$ violation, to arbitrary precision.

However, in the numerical scan to be described in the next section we will not employ this reverse engineering algorithm
but rather stick to the standard procedure. While the reverse engineering algorithm is useful to prove the existence of a solution fitting the masses and mixing, it can very easily fall in a small region of the allowed parameters space. Since we will be interested in fitting additional observables, we must try to explore the largest region of the parameters space possible. 
We observed that the standard $\chi^2$ minimization procedure, starting with a seed point, quickly
converges to a nearby parameter space point which reproduces quark masses, mixing, and CPV arbitrarily well.
The largest amount of the computer time in the scan was spent to find points which fit sufficiently well
the meson oscillation observables.

The Yukawa matrices in cases $B_1$ \eqref{caseB1} and $B_2$ \eqref{caseB2} have 7 complex and one real free parameters,
making in total 15 real free parameters in each sector. In case $B_3$, we have in total 13 real free parameters per sector.
We remind that in the SM, when we start from arbitrary complex matrices $\Gamma$ and $\Delta$,
we have 18 initial free parameters in each sector.
Certainly many of these free parameters are superficial and can be removed by basis changes in the left-handed and right-handed
quark sectors. In CP4 3HDM, we have {\em fewer} superficial free parameters.
For example, rotations of the right-handed quark fields do modify some observables, namely, 
the off-diagonal FCNC elements linked to extra neutral Higgs fields. However, these elements are, themselves, correlated,
which renders counting the independent physical parameters not a straightforward exercise. 
We postpone this calculation to a future work.

\subsection{Flavour-changing neutral currents}

The generic presence of Higgs-exchange-induced large FCNCs at tree level is a notorious problem of all multi-Higgs-doublet
models. For example, in the down quark sector, we get after electroweak symmetry breaking
\be
\bar d_L \left[M_d + {1\over \sqrt{2}} \sum_i  \Gamma_i (h_i + i a_i) \right] d_R + h.c.
\ee
When diagonalizing the quark mass matrix by a bi-unitary transformation $d_L = V_{dL} d_L^{ph.}$, $d_R = V_{dR} d_R^{ph.}$,
\be
V_{dL}^\dagger M_d V_{dR} = D_d = \mathrm{diag}(m_d, m_s, m_b)\,,
\ee
one does not automatically diagonalize the individual Yukawa matrices $\Gamma_i$ which describe coupling of the scalar fields to
the physical quarks leading to tree-level FCNCs. This can be conveniently described in the Higgs basis, in which the Yukawa
matrices $\Gamma_i^{(H)}$ are expressed via $\Gamma_i$ of the real vev basis as
\bea
\Gamma_1^{(H)} &=& {\sqrt{2} \over v} M_d\,,\nonumber\\
\Gamma_2^{(H)} &=& - s_\psi \Gamma_2 + c_\psi \Gamma_3\,,\nonumber\\
\Gamma_3^{(H)} &=& s_\beta \Gamma_1 - c_\beta(\Gamma_2 c_\psi + \Gamma_3 s_\psi)
= - \cot\beta {\sqrt{2} \over v} M_d + {1 \over s_\beta} \Gamma_1\,.\label{FCNCmatrices}
\eea
The Yukawa interactions can be written as
\bea
&&\bar d_L \left(\sum_j \Gamma_j^{(H)} \Phi_j\right) d_R + h.c. =
\bar d_L \left[M_d + {1 \over \sqrt{2}}\left(\Gamma_1^{(H)} \rho_1 + \Gamma_2^{(H)} \Phi_2
+ \Gamma_3^{(H)} \Phi_3 \right)\right]d_R + h.c.\nonumber\\
&=&
\bar{d_L}^{ph.} D_d {d_R}^{ph.}\left(1 + {\rho_1 \over v}\right)
+ {1 \over\sqrt{2}}\bar{d_L}^{ph.} \left[ \Gamma_2^{(H, ph.)} \Phi_2 + \Gamma_3^{(H, ph.)} \Phi_3\right]{d_R}^{ph.}+ h.c.
\label{Yukawa-FCNCs}
\eea
In the last line, we switched to physical quarks $d^{ph.}$ and introduced the following matrices
\be
\Gamma_{2,3}^{(H, ph.)} \equiv V_{dL}^\dagger \Gamma_{2,3}^{(H)} V_{dR}\,.\label{FCNCmatrices-phys}
\ee
The key observation is that the CP4-symmetric Yukawa structures $\Gamma_i$ generically produce matrices
$\Gamma_{2,3}^{(H, ph.)}$ with unsuppressed off-diagonal terms, typically of the same order as the off-diagonal
elements of $\Gamma_{2,3}$. Even if some of these off-diagonal elements are small due to accidental cancellations
assisted by vevs, this cannot happen with all off-diagonal elements, again due to the specific structure of
the CP4 3HDM Yukawa sector. Notice also that one cannot assume that entries $g_{ij}$ are very small in
$\Gamma_{2,3}$ since those, when multiplied by $v_2$ and $v_3$, must provide sufficient
mass splitting for quarks.

Thus, the only solution in sight is to assume that a strong or exact alignment takes place in the scalar sector,
as discussed in Section~\ref{section-scalar-alignment}. In this case, the field $\rho_1$ in Eq.~\eqref{Yukawa-FCNCs}
is identified with the 125 GeV Higgs boson $H_{125}$ and it is protected from tree-level FCNCs. However,
dangerous FCNCs arise from the other neutral Higgs bosons' exchanges, unless they are sufficiently heavy or other
cancellations take place. One then faces the necessity of a numerical study in order to check if the resulting
models can pass the tight experimental constraints.


\section{Numerical scan}
\label{section:scan}

In this Section, we shall describe the procedure of scanning of the parameter space of the model in the Higgs sector (scalar potential) and in the Yukawa sector, 
with different combinations of Yukawa matrices $\Gamma_a$ and $\Delta_a$ discussed above. Our intent is to show that the CP4-3HDM scenarios 
can realistically fit the current data avoiding large departures from the SM predictions in well-measured observables.

The Yukawa sector scan will take as input the vevs and the diagonalizing rotation matrices of the scalar fields. Therefore, we first analyze the scalar 
sector which, at least at tree level, requires no information from the fermionic sector. Then, we perform an efficient numerical scan over the Yukawa 
sector parameters. At each run, we start from a seed point, converge to a point in the Yukawa parameter space that gives a good description of quark masses 
and the CKM matrix, and then check $K$-meson parameters $\epsilon_K$ and $\Delta m_K$. Keeping only those points which are sufficiently close to their 
experimental values, we proceed to the $B$-meson oscillation parameters. At the end, we have a fair selection of good points, which pass all the criteria 
we have imposed. Let us describe the numerical procedure as well as the main results in detail.

\subsection{Parameter space of the scalar sector}

As we have have seen in Section~\ref{section-scalar-alignment}, we can guarantee the presence of a SM-like Higgs field $H_{125}$
by working in the scalar alignment limit. In the present analysis, we shall take this limit and explore the corresponding predictions.

\begin{figure}[h]
\begin{tabular}{c}
\includegraphics[width=01\textwidth]{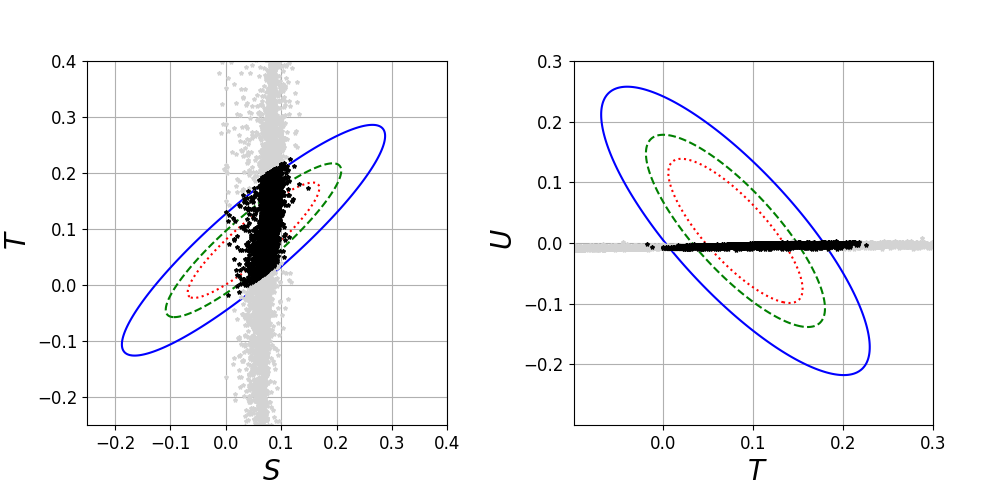}
\end{tabular}
\caption{\label{Fig:STU} Oblique parameters $STU$. Left: $T$ vs. $S$ with the ellipses drawn for $90\%$, $95\%$ and $99\%$ confidence level with 
$U$ fixed at its the best fit value. Right: $U$ vs. $T$ with the ellipses drawn for $90\%$, $95\%$ and $99\%$ confidence level with $S$ fixed at 
its the best fit value.}
\end{figure}

The basic procedure adopted in our exploration of the parameter space can be summarized as follows:
\begin{itemize}

\item The alignment limit, i.e. $m_{11}^2=m_{22}^2$, has been imposed in the mass matrices~\eqref{mass-matrix-charged} and~\eqref{neutralMM}. 
It can be formulated as
\begin{align}
\begin{split}
\text{Re}(\lambda_9)=&\frac{1}{t_{2\psi}}\left[\lambda_{34}-2\lambda_2+\lambda_5s_{2\psi}+\frac{v_1^2}{u^2}(2\lambda_1-\lambda_{34})\right]\,.
\end{split}
\end{align}
In this limit, as shown in Section~\ref{section-scalar-alignment}, the SM-like Higgs mass has a simple analytic expression, 
Eq.~\eqref{alignment-condition-SM-mass}. We can then use this expression to eliminate one extra parameter, i.e.
\begin{equation}
\lambda_1={1 \over 2 v_1^2} \left[m_{H_{125}}^2-u^2\left(\lambda_{34}+\lambda_5 s_{2\psi}\right)\right]\,.
\end{equation}

\item At this stage we have 9 free parameters: $\lambda_{2,3,4,5}$, $\lambda^{\prime}_{3,4}$ $\text{Im}(\lambda_8)$, and two vevs ratios,
which can be expressed in terms of angles $\psi$ and $\beta$. We choose an initial random point satisfying the BFB conditions, presented in 
Eq.~\eqref{copositivity=3x3}. In order to stay in the perturbative regime we took\footnote{
Thanks to the results of Ref.~\cite{Bento:2017eti}, there exist now the exact expressions
for the perturbativity constraints in CP4 3HDM. Since they involve numerically solving
polynomial equations and since we only look for examples of benchmark points and do not aim 
to provide the exhaustive parameter space scan, we did not implement them. 
We believe that our simplified constraints are on the conservative side
and we do not expect the exact unitarity constraints to significantly modify our results.}
\begin{equation}
|\lambda_i^{(\prime)}|,\ |\text{Re}(\lambda_{8,9})|,\ |\text{Im}(\lambda_{8,9})|\leq 5\,.
\end{equation}
For the vevs we allowed the ratios $u/v_1 = t_\beta$ and $v_3/v_2=t_\psi$ to be between 0 and 100.

\item We now use this point to compute the mass eigenstates and the respective field rotations. We use \textsf{iminuit 1.2}~\cite{iminuit} in order 
to find the values of the free parameters that minimize the function
\begin{equation}
\chi_{\text{scalar}}^2=\frac{(m_{\text{lightest}}^2-(125\,\rm{GeV})^2)^2}{\sigma_{\rm Higgs}^2}\,,
\end{equation}
where $m_{\text{lightest}}$ is the numerical value of the mass for the lightest massive neutral scalar, and 
$\sigma_{\rm Higgs}$ is the allowed standard deviation. Since one of the mass eigenstates is $H_{125}$, whose mass by construction 
is set to 125 GeV, this $\chi_{\text{scalar}}^2$-minimization promotes the lightest massive neutral state to be the observed SM-like Higgs boson.
We then take an extremely small deviation $\sigma_{\rm Higgs}=10^{-10}$ GeV$^2$, 
such that the method can quickly converge to the scenario where 
the massive state that is aligned with the Higgs basis is the lightest massive neutral scalar with mass of $125$ GeV. Guaranteeing the existence 
of this state automatically ensures that the neutral scalar Hessian matrix is non-negative. We also make sure that the resulting points of 
the scalar parameter space produce positive mass squared for the charged Higgs bosons.

The assumption that the 125 GeV Higgs is the lightest one is not obligatory. Whether CP4 3HDM can accommodate lighter
Higgs states and not run into an immediate conflict with the experimental data deserves a dedicated study, which we delegate to a future work.

\item Once the minimization of $\chi_{\rm scalar}^2$ is achieved, we check if the final values of the scalar potential couplings satisfy the BFB conditions. 
Also, as a safety check, we verify if all the fields' masses squared are positive and if the lightest neutral massive one is actually the SM-like Higgs boson.

\end{itemize}
The presence of additional neutral and charged scalars can lead to large deviations in the well measured electroweak observables.
The most constraining electroweak precision observables can be summarized into the three oblique parameters $S, T, U$~\cite{Kennedy:1988sn,
Peskin:1990zt,Maksymyk:1993zm}.
We have computed these parameters, as well as three additional ones $V$, $W$ and $X$ (with no relevant impact on the parameter space), using the corresponding 
expressions found in Ref.~\cite{Grimus:2008nb}. The current electroweak precision measurements lead to~\cite{Patrignani:2016xqp}
\begin{equation}
S=0.05\pm 0.10\,,\quad T=0.08\pm 0.12\,,\quad U=0.02\pm 0.10\,.
\end{equation}
These parameters are strongly correlated: there is a $91\%$ correlation between $S$ and $T$ parameters, while the $U$ parameter is
$-61\%$ and $-82\%$ anti-correlated with $S$ and $T$, respectively. In Fig.~\ref{Fig:STU} we present the $TS$- and $UT$-plane where the third 
oblique parameter is set to its best fit value. We can see that both $S$ and $U$ place no constraints on the parameter space, while $T$ restricts the allowed 
parameters space significantly. The points surviving the $3\sigma$ constraints from the $STU$ oblique parameters are represented by black dots while 
the remaining points are shown by gray dots. This representation will be used throughout this section.

\begin{figure}[h]
\begin{center}
\includegraphics[width=0.55\textwidth]{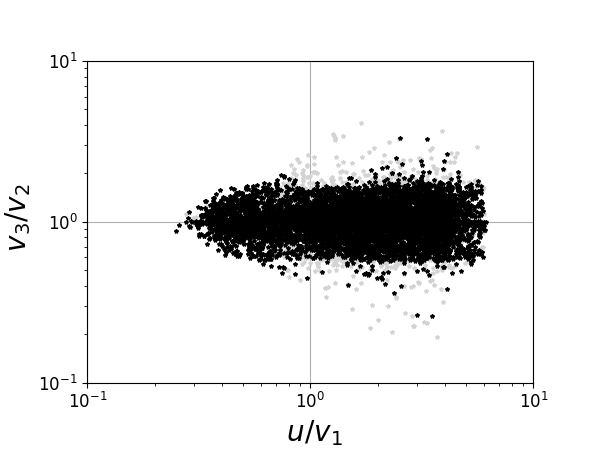}
\end{center}
\caption{\label{Fig:vevs} Allowed region for the vev ratios $v_3/v_2$ and $u/v_1$.}
\end{figure}

We can now explore some trends for the vevs and the scalar spectrum which emerge in the CP4 3HDM scan. 
In Fig.~\ref{Fig:vevs} we show the region of the vev ratios for the points which pass the scalar sector constraints 
discussed above. Both ratios tend to be $\mathcal{O}(1)$, even though we initially allowed for much wider ranges. 
The shape of this region arises from an interplay of several factors:
the BFB and perturbativity constraints, the scalar alignment assumption Eq.~(\ref{alignment-condition}), 
the requirement that $H_{125}$ be the lightest massive neutral scalar, 
and finally the relation among the vev ratios and $\lambda$'s encoded in Eq.~(\ref{t23}).
Qualitatively, choosing the vev ratios very far from unity would push some of the quartic couplings
beyond the allowed range.
If the scalar alignment requirement or the constraint on the lightest massive neutral scalar were dropped,
larger regions on this plot would be accessible.  

\begin{figure}[h]
\begin{tabular}{c}
\includegraphics[width=1\textwidth]{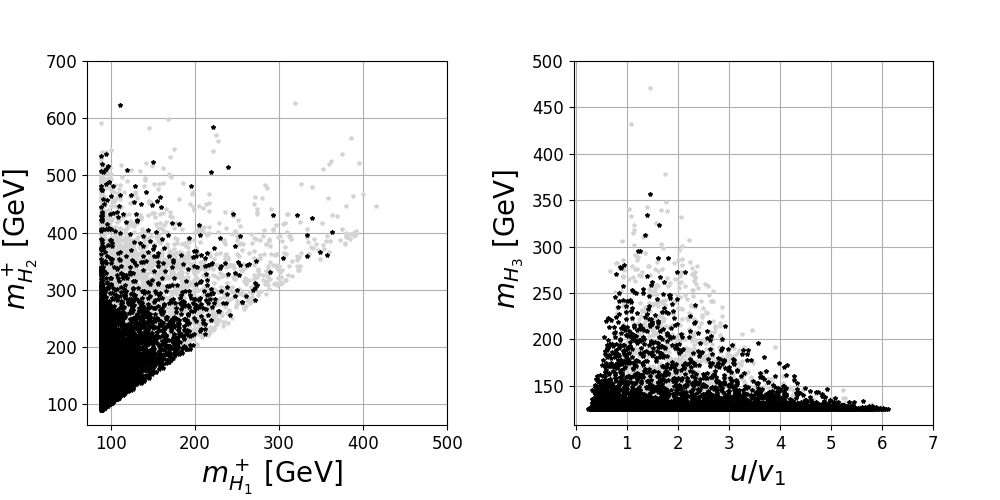}
\end{tabular}
\caption{\label{Fig:MassSpec1} Left: Charged scalar spectrum $m_{H_2}^+$ vs. $m_{H_1}^+$. Right: next-to-lightest neutral scalar $H_3$ 
mass dependence with respect to $u/v_1$.}
\end{figure}

In Fig.~\ref{Fig:MassSpec1} (left) we present the masses of the two charged scalars $H_1^+$ and $H_2^+$. One sees that the mass of 
lightest charged scalar can go down to as low as $90$ GeV and still be compatible with the electroweak precision data.
\begin{figure}[h]
\begin{tabular}{c}
\includegraphics[width=0.33\textwidth]{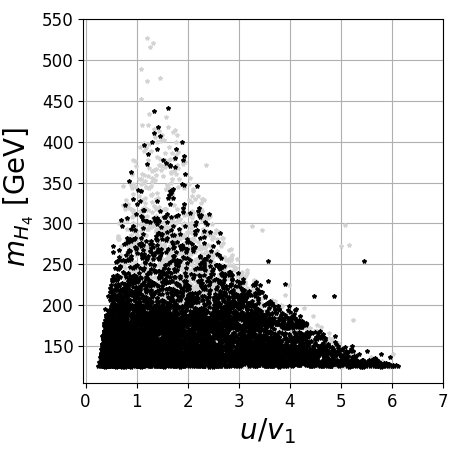}
\includegraphics[width=0.33\textwidth]{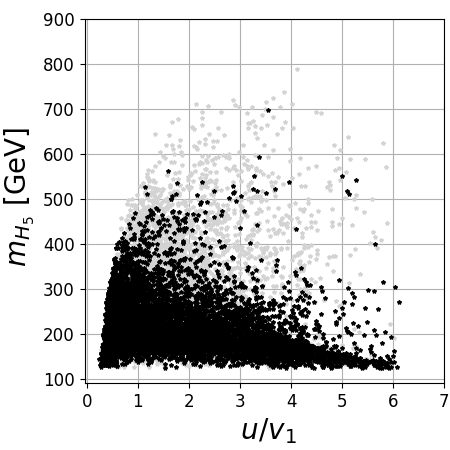}
\includegraphics[width=0.33\textwidth]{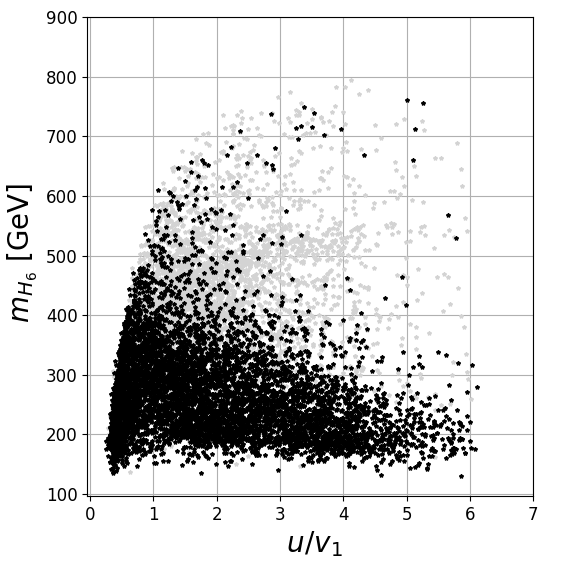}
\end{tabular}
\caption{\label{Fig:MassSpec2} The masses of heavier neutral Higgses $H_4$, $H_5$, and $H_6$ vs. the vev ratio $u/v_1$.}
\end{figure}
We label the five neutral physical scalars as in Appendix~\ref{appendix-three-bases}:
\be
H_{125},\ H_3,\ H_4,\ H_5,\ H_6\,.\label{physical-neutrals}
\ee
These states are mass ordered. In Fig.~\ref{Fig:MassSpec1}, right, 
we plot the next-to-lightest neutral scalar mass $m_{H_3}$ vs. the ratio of two vevs.
Again, the same constraints that were shaping Fig.~\ref{Fig:vevs} are at work here.
In addition, one can directly compute the sum of all neutral masses squared as 
$\sum_i m_{H_i}^2 = \Tr{{\cal M}_n}$. The trace of the neutral scalar mass matrix (\ref{neutralMM})
is given in terms of $v^2$, vev ratios, and quartic couplings, and therefore it is bounded from above. 
Satisfying all scalar sector constraints maximizes $\sum_i m_{H_i}^2$ when all vevs are of the same order.
Since the scalars (\ref{physical-neutrals}) are mass-ordered, the largest value of $m_{H_3}$ is attained
when all additional neutral Higgses are equally heavy, which explains the rather sharp upper boundary 
of Fig.~\ref{Fig:MassSpec1}, right. 

The mass spectra of yet heavier Higgses $H_4$, $H_5$, $H_6$ are shown in Fig.~\ref{Fig:MassSpec2}.
Here, the upper boundaries pass higher and are less pronounced. For example, maximizing $M_{H_5}$ implies
keeping $H_3$ and $H_4$ close to $H_{125}$ and setting $M_{H_5} = M_{H_6}$.

\subsection{Parameter space of the Yukawa sector}

Once the scalar sector analysis is done, we have a numerical set of vevs and scalar rotation matrices, which can be used as input for the Yukawa sector scan.
In this sector, we are interested in finding the parameter space points where the experimentally known masses of the quarks as well as the CKM mixing matrix 
can be fully accommodated. We used the following experimental data~\cite{Patrignani:2016xqp}:
\begin{align}
\begin{split}
&m_d=4.7\pm 0.5 \ \text{MeV}\,,\quad m_s=96\pm 8\ \text{MeV}\,,\quad m_b=4.18\pm 0.04\  \text{GeV}\\
&m_u=2.2\pm 0.6\ \text{MeV}\,,\quad m_c=1.28\pm 0.03\ \text{GeV}\,,\quad m_t=173.1\pm 0.6 \ \text{GeV}
\end{split}
\end{align}
for the quark masses,
\begin{equation}
\sin\theta_{12}=0.22496\pm 0.001\,,\quad \sin\theta_{23}=0.0416496\pm 0.001\,,\quad \sin\theta_{13}=0.00361726\pm 0.0001
\end{equation}
for the CKM mixing angles, and
\begin{equation}
\sin\delta = 0.949\pm 0.01
\end{equation}
for the $CP$ violating phase.

On top of fitting the masses and mixings in the quark sector, we also need to guarantee that we do not have too large FCNCs. The meson oscillation observables 
set strong constraints on these models. For the $K$-meson sector, we have the CPV observable $\epsilon_K$ and the mass difference $\Delta m_K$ between 
$K_L$ and $K_S$. The observables read~\cite{Buras:2013rqa}
\begin{equation}
\epsilon_{K}=\frac{\kappa_\epsilon e^{i\varphi_\epsilon}}{\sqrt{2}(\Delta m_K)_{\text{exp}}}\left[\text{Im}(M_{12}^K)\right]\,,\quad \Delta m_K =2\text{Re}(M_{12}^K)\,,
\end{equation}
with
\begin{equation}
\left(M_{12}^K\right)^\ast=\frac{G_F^2}{12\pi^2}F_K^2\hat{B}_K m_K m_W^2
\left[\lambda_c^2 \eta_1 x_c+\lambda_t^2\eta_2 (S_0(x_t)+\Delta S(K))+
2\lambda_c\lambda_t \eta_3 S_0(x_c,x_t)\right] \,.
\end{equation}
Here, $F_K=156.1$ MeV is the kaon decay constant, $\hat{B}_K=0.767$ accounts for SM non-perturbative corrections, the factors $\eta_i$ encode the QCD corrections 
and take the values $\eta_1=1.87$, $\eta_2=0.5765$ and $\eta_3=0.496$. The coefficients $\lambda_i=V_{is}^\ast V_{id}$, where $V$ is the CKM mixing matrix. 
The loop functions $S_0(x)$ and $S_0(x,y)$ are given in Ref.~\cite{Blanke:2006sb} (see also Ref.~\cite{Buras:2013rqa}). Finally, $\varphi_\epsilon=43.51^\circ$ 
and $\kappa_\epsilon=0.94$ were taken from Ref.~\cite{Patrignani:2016xqp,Buras:2008nn,Buras:2010pza}.

In the above expression, the function $\Delta S(K)$ is the only model dependent contribution. For our model, the contribution of the neutral scalar fields is given by
\begin{equation}\label{eq: DeltaSK}
\Delta S(K)= -\sum_{i=2}^{6}\left[\frac{\left(\Delta_L^{sd}(H_i)\right)^2}{2m_{H_i}^2}f_{LL}(\mu)+\frac{\left(\Delta_R^{sd}(H_i)\right)^2}{2m_{H_i}^2}f_{RR}(\mu)+
\frac{\Delta_L^{sd}(H_i)\Delta_R^{sd}(H_i)}{2m_{H_i}^2}f_{LR}(\mu)\right]\,,
\end{equation}
where the functions $f_{XY}(\mu_H)$ encode the information on the Wilson coefficients and hadronic matrix elements. Their explicit form, as well as numerical values, 
can be found in Ref.~\cite{Buras:2013rqa}. The relevant part here is the model dependent $\Delta_X^{sd}(H_i)$, which in our model reads
\begin{equation}
\Delta_L^{\alpha\beta}(H_i)^\dagger=\Delta_R^{\alpha\beta}(H_i)=\frac{1}{\sqrt{2}}\sum_{j=1}^3\left(\Gamma_j\right)_{\alpha\beta}(R_{ij}+i R_{i\,3+j})\,.
\end{equation}
We have included in the $\chi^2$-fit the $K$-meson sector, with the experimental values for the observables~\cite{Patrignani:2016xqp}
\begin{equation}
\left|\epsilon_{K}\right|^{exp}=2.228\times 10^{-3}\,,\quad \Delta m_K^{exp}=3.5\times 10^{-15}\,\text{GeV}\,,
\end{equation}
and we allowed a $50\%$ deviation from the experimental measurement.

The scan procedure is very similar to the one adopted in the scalar sector, with the $\chi^2$ function to be now minimized defined as
\begin{equation}
\chi^2_{yuk.}=\frac{1}{12}\left[\sum_{i=1}^{6}\frac{(m_i^{exp}-m_i)^2}{\sigma_{m_i}^2}+\sum_{i=1}^4\frac{(s_{\theta_i}^{exp}-s_{\theta_i})^2}{\sigma_{\theta_i}^2}+
\frac{(|\epsilon_K|^{exp}-\epsilon_K)^2}{\sigma_\epsilon^2}+\frac{(\Delta m_K^{exp}-\Delta m_K)^2}{\sigma_{\Delta m}^2}\right] \,.
\label{chi2-yuk}
\end{equation}
Here, $m_i$ are the quark masses, $s_{\theta_i}$ correspond to the sine of mixing angles and CPV phase. In the numerical scan, the absolute values of the Yukawa parameters 
were taken in the range $[0,5]$ with arbitrary phases. Once the numerical quark mass matrices are found, we extract the mixing angles and the CPV phase through 
the reshaping invariants
\begin{align}
\begin{split}
&s_{12}=\frac{|V_{12}|}{\sqrt{|V_{11}|^2+|V_{12}|^2}} \,,\, s_{23}=\frac{|V_{23}|}{\sqrt{|V_{11}|^2+|V_{12}|^2}} \,,\, s_{13}=|V_{13}|\,,\,  
s_\delta=8\frac{\text{Im}(V_{11}V_{22}V_{12}^\ast V_{21}^\ast)}{c_{13}s_{2\theta_{12}}s_{2\theta_{13}}s_{2\theta_{23}}}\,.
\end{split}
\end{align}
\begin{figure}[h]
\begin{center}
\includegraphics[width=1\textwidth]{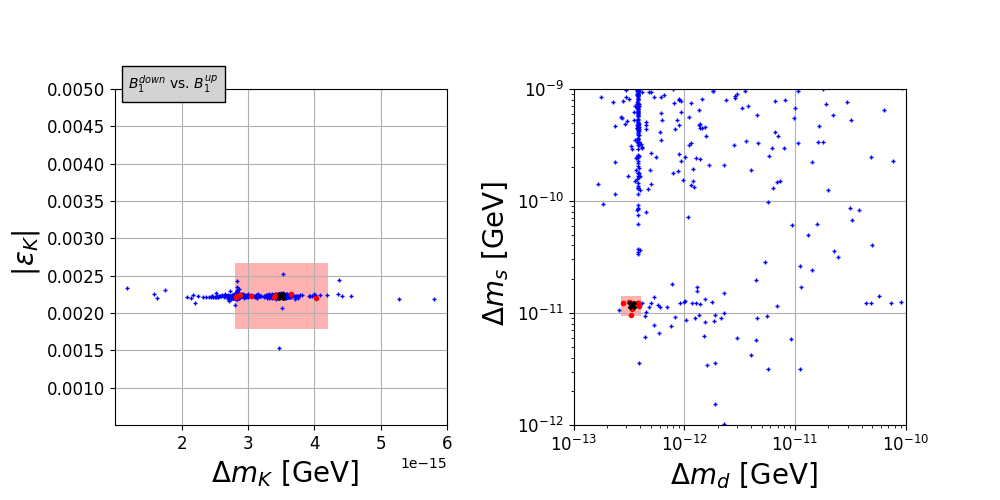}\\
\includegraphics[width=1\textwidth]{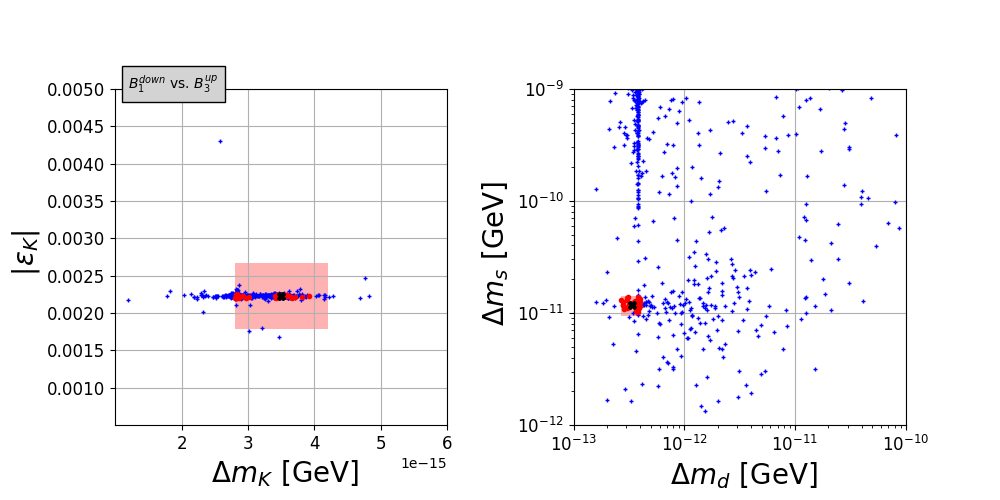}
\end{center}
\caption{\label{Fig:MesonB1} Meson sector predictions for $(B_1, B_1)$ (upper) and $(B_1, B_3)$ (lower) scenarios.
The red dots indicate the points, which satisfy the $K$- and $B$-meson oscillations parameters within the chosen margins shown 
by pink rectangles.}
\end{figure}
\begin{figure}[h]
\begin{center}
\includegraphics[width=1\textwidth]{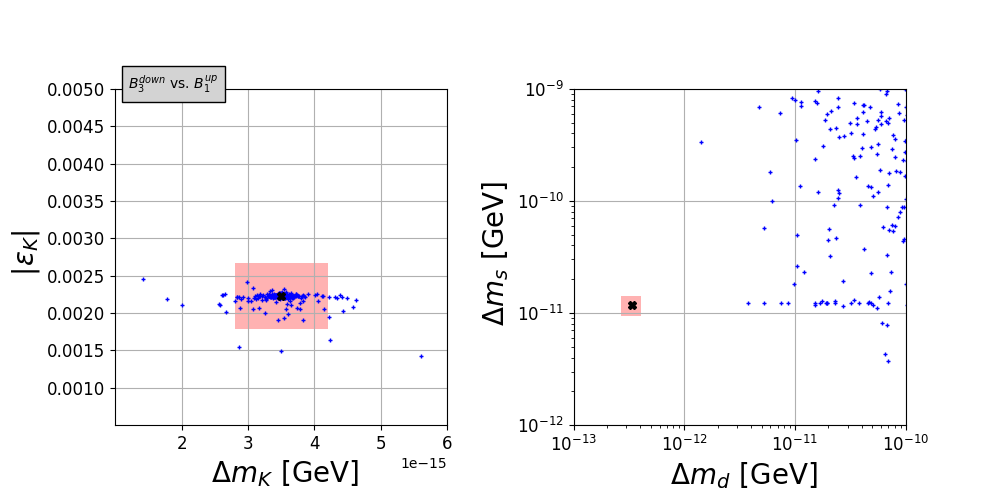}\\
\includegraphics[width=1\textwidth]{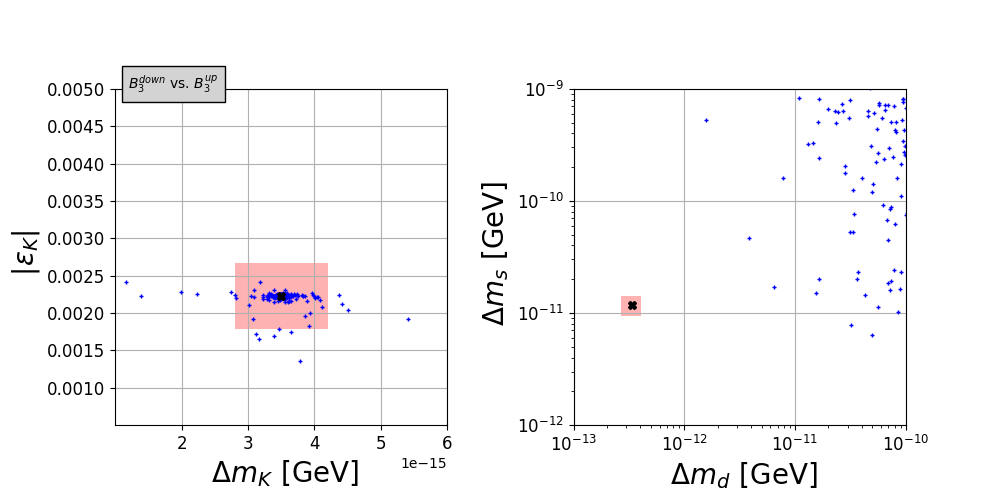}
\end{center}
\caption{\label{Fig:MesonB3} Meson sector predictions for $(B_3, B_1)$ (upper) and $(B_3, B_3)$ (lower) scenarios.}
\end{figure}
\begin{figure}[h]
\begin{center}
\includegraphics[width=1\textwidth]{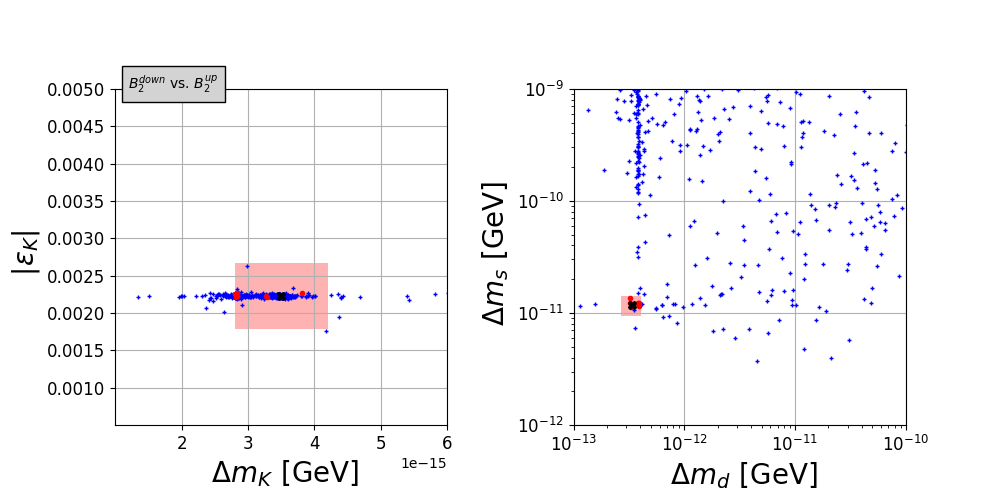}\\
\includegraphics[width=1\textwidth]{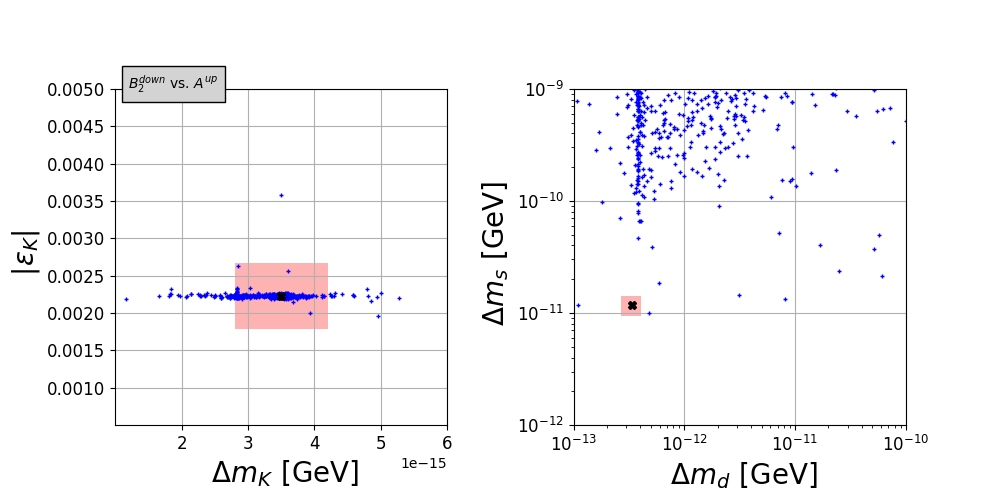}
\end{center}
\caption{\label{Fig:MesonB2} Meson sector predictions for $(B_2, B_2)$ (upper) and $(B_2, A)$ (lower) scenarios.}
\end{figure}
For the $B$-meson sector, the mass difference can also set strong constraints on the parameter space. From Ref.~\cite{Buras:2013rqa} we have
\begin{equation}
\Delta m_{d,s}=\frac{G_F^2}{6\pi^2}m_W^2m_{B_{d,s}}|\lambda_t^{(d,s)}|^2 F_{B_{d,s}}^2 \hat{B}_{d,s} \eta_B \left|S_0(x_t)+\Delta S(B_{d,s})\right|\,,
\end{equation}
with the decay constants $F_{B_d}=188$ MeV and $F_{B_s}=225$ MeV, the meson masses $m_{B_d}=5279$ MeV and $m_{B_s}=5366$ MeV, 
and the factors $\hat{B}_d=1.26$, $B_{s}=1.33$ and $\eta_B=0.55$. Finally, the CKM factor is $\lambda_t^{(q)}=V^\ast_{tb}V_{tq}$.

The function $\Delta S(B_{d,s})$ takes a form similar to Eq.~\eqref{eq: DeltaSK}, where the functions $f_{XY}(\mu)$ have distinct values and 
$\Delta_X^{sd}(H_i)$ is replaced by $\Delta_X^{b(s,d)}(H_i)$. The current measurements give the following best fit values~\cite{Abulencia:2006ze,
Aaij:2011qx,Amhis:2016xyh}:
\begin{equation}
\Delta m_d^{exp}= 3.37\times  10^{-13}\, \text{GeV}\,,\quad \Delta m_s^{exp}= 1.17\times 10^{-11}\, \text{GeV}\,.
\end{equation}
In Figs.~\ref{Fig:MesonB1} to~\ref{Fig:MesonB2} we show the predictions on $\Delta m_d$ and $\Delta m_s$ for the Yukawa sector combinations
available in the CP4 3HDM. The pink rectangles delimit the regions with $20\%$ deviation from the experimental measurement, while the red dots 
are points in the parameter space that are in agreement with both $K$ and $B-$meson oscillations. We see that out of the 7 down-up model 
implementations, only 4 survive the meson oscillation constraints. The surviving models are $(B_1, B_1)$, $(B_1, B_3)$, $(B_2, B_2)$, 
and the model with no FCNCs in the down-quark sector $(A, B_2)$.
\begin{figure}[h]
\begin{center}
\includegraphics[width=1.0\textwidth]{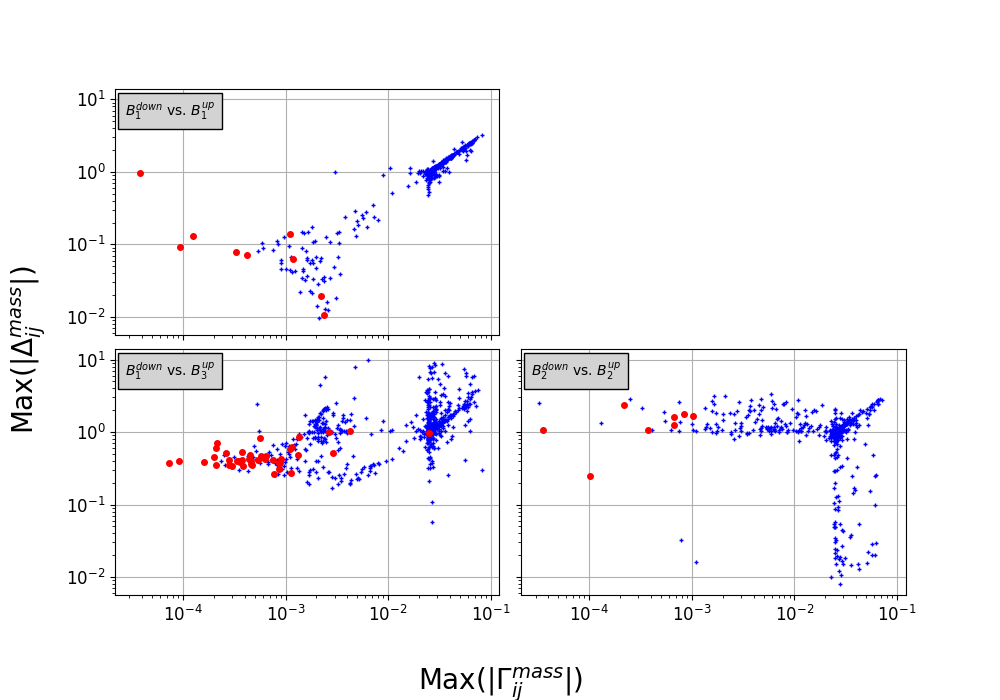}
\end{center}
\caption{\label{Fig:FCNCs} Maximal values of off-diagonal elements in the mass basis for the scenarios
$(B_1, B_1)$ (upper left), $(B_1, B_3)$ (lower left), and $(B_2, B_2)$ (lower right).}
\end{figure}

For the three models surviving the meson oscillations constraints despite non-trivial FCNCs in the down-quark sector, 
we looked at the magnitude of the off-diagonal Yukawa entries in the mass basis. 
These are shown in Fig.~\ref{Fig:FCNCs}. 
Namely, we computed the Yukawa matrices in the Higgs basis and for quark mass eigenstates $\Gamma_{2,3}^{(H, ph.)}$,
which were defined in Eq.~\ref{FCNCmatrices-phys}, and extracted the off-diagonal entry in either of these two matrices
with largest absolute value, denoted as Max$(|\Gamma_{ij}^{mass}|)$.
We repeated the same for the up-quark sector extracting largest off-diagonal element from $\Delta_{2,3}^{(H, ph.)}$,
denoted as Max$(|\Delta_{ij}^{mass}|)$.
Then, for each point of the scan, we plotted these two values to get the scattering plots Fig.~\ref{Fig:FCNCs}.
These plots exhibit certain clustering around $\sqrt{2}m_b/v$ and $\sqrt{2}m_t/v$ for down and up sectors
which is natural when FCNCs are unsuppressed. Most of these points, however, do not pass the tight meson oscillation
constraints. Those which do are shown in red.
For the down-quark sector, the largest absolute value for the off-diagonal entry 
typically stay below $10^{-3}$, although a few outliers exist.
In the up-quark sector, the off-diagonal elements can indeed be large.

What emerges from this analysis is that the magnitudes of the off-diagonal elements 
do not necessarily need to be extremely small to provide a good description of the data. 
Several neutral Higgs bosons, which mediate FCNCs, can interfere destructively due to the intrinsic 
structural properties of the model, and lead to acceptable results. This is also reminiscent of what was observed in certain classes
of the 2HDM, where the real and imaginary parts of the complex neutral Higgs field destructively interfere
and bring FCNCs under control, see e.g.~Ref.~\cite{Botella:2014ska}.

In order to illustrate the typical textures which arise in this scan, we give in Table~\ref{Tab: BP} all the parameters for two benchmark points 
which correspond to cases $(B_1^{\rm down}, B_1^{\rm up})$ and $(B_1^{\rm down}, B_3^{\rm up})$. 
In spite of the presence of several Higgs bosons which are only moderately heavy,
the FCNC-induced contributions to the kaon and $B$-meson mixing are under control. 
In both cases, the value of $\chi_{yuk.}^2$ given by (\ref{chi2-yuk}) is very small, about $0.01$,
which indicates that the model is capable of describing the quark masses, mixing, and CPV to arbitrary precision, 
as discussed in Section~\ref{section-reconstructing-yukawas}.
\begin{table}[!ht]
\begin{center}
\begin{tabular}{|ll|ll|}
\hline
\multicolumn{4}{|c|}{{\bf case $(B_1, B_1)$}}\\[1mm]
\hline\hline
\multicolumn{2}{|c|}{down-sector}&\multicolumn{2}{|c|}{up-sector}\\
$|g_{11}|=4.848\times 10^{-4}$ & $\text{Arg}(g_{11})=2.139$ & $|g_{11}|=1.375\times 10^{-4}$ & $\text{Arg}(g_{11})=5.780$\\
$|g_{12}|=1.362\times 10^{-4}$ & $\text{Arg}(g_{12})=2.754$ & $|g_{12}|=3.888\times 10^{-5}$ & $\text{Arg}(g_{12})=1.726$\\
$|g_{13}|=2.402\times 10^{-5}$ & $\text{Arg}(g_{13})=1.544$ & $|g_{13}|=0.00202$ & $\text{Arg}(g_{13})=3.688$ \\
$|g_{21}|=1.854 \times 10^{-4}$ & $\text{Arg}(g_{21})=3.632$ & $|g_{21}|=0.00109$ & $\text{Arg}(g_{21})=6.242$  \\
$|g_{22}|=3.504 \times 10^{-4}$ & $\text{Arg}(g_{22})=1.102$ & $|g_{22}|=2.597\times 10^{-4}$ & $\text{Arg}(g_{22})=2.888$  \\
$|g_{23}|=2.705 \times 10^{-6}$ & $\text{Arg}(g_{23})=0.000$ & $|g_{23}|=0.04947$ & $\text{Arg}(g_{23})=5.465$  \\
$|g_{31}|=0.00275$ & $\text{Arg}(g_{31})=5.402$ & $|g_{31}|=0.23005$ & $\text{Arg}(g_{31})=5.373$ \\
\multicolumn{2}{|c|}{$g_{33}=0.04735$} & \multicolumn{2}{|c|}{$g_{33}=1.93847$}\\
\hline\hline
\multicolumn{4}{|c|}{$v_1 = 88.0$ GeV, $\quad v_2 = 121.4$ GeV, $\quad v_3 = 88.5$ GeV}\\
\multicolumn{4}{|c|}{$m_{H_3} = 171.3$ GeV, $\quad m_{H_4} = 342.4$ GeV, $\quad m_{H_5} = 345.9$ GeV, $\quad m_{H_6} = 395.5$ GeV}\\
\multicolumn{4}{|c|}{$m_{H_1^+} = 188.6$ GeV, $\quad m_{H_2^+} = 248.4$ GeV}\\
\multicolumn{4}{|c|}{$|\epsilon_K|= 2.222\times 10^{-3}$, $\quad \Delta m_K = 2.833 \times 10^{-15}$ GeV}\\ 
\multicolumn{4}{|c|}{$\Delta m_{B_d} = 3.823\times 10^{-13}$ GeV, $\quad \Delta m_{B_s} = 1.227\times 10^{-11}$ GeV}\\
\hline
\end{tabular}
\smallskip
\begin{tabular}{|ll|ll|}
\hline
\multicolumn{4}{|c|}{{\bf case $(B_1, B_3)$}}\\[1mm]
\hline\hline
\multicolumn{2}{|c|}{down-sector}&\multicolumn{2}{|c|}{up-sector}\\
$|g_{11}|=5.565\times 10^{-4}$ & $\text{Arg}(g_{11})=2.792$ & $|g_{11}|=0.00118$ & $\text{Arg}(g_{11})=6.030$\\
$|g_{12}|=2.839\times 10^{-4}$ & $\text{Arg}(g_{12})=0.039$ & $|g_{12}|=0.00158$ & $\text{Arg}(g_{12})=3.136$\\
$|g_{13}|=3.243\times 10^{-4}$ & $\text{Arg}(g_{13})=0.150$ & $|g_{13}|=0.11888$ & $\text{Arg}(g_{13})=5.901$ \\
$|g_{21}|=3.547 \times 10^{-4}$ & $\text{Arg}(g_{21})=2.735$ && \\
$|g_{22}|=1.322 \times 10^{-4}$ & $\text{Arg}(g_{22})=0.472$ && \\
$|g_{23}|=5.987\times 10^{-4}$ & $\text{Arg}(g_{23})=3.335$ & $|g_{23}|=0.12324$ & $\text{Arg}(g_{23})=3.077$ \\
$|g_{31}|=0.02066$ & $\text{Arg}(g_{31})=2.855$ & $|g_{31}|=0.20278$ & $\text{Arg}(g_{31})=0.419$ \\
  & & $|g_{32}|=0.21372$ & $\text{Arg}(g_{32})=0.019$\\
\multicolumn{2}{|c|}{$g_{33}=0.00196$} & \multicolumn{2}{|c|}{$g_{33}=-1.19372$}\\
\hline\hline
\multicolumn{4}{|c|}{$v_1 = 142.8$ GeV, $\quad v_2 = 66.1$ GeV, $\quad v_3 = 74.6$ GeV}\\
\multicolumn{4}{|c|}{$m_{H_3} = 220.4$ GeV, $\quad m_{H_4} = 304.4$ GeV, $\quad m_{H_5} = 318.9$ GeV, $\quad m_{H_6} = 352.2$ GeV}\\
\multicolumn{4}{|c|}{$m_{H_1^+} = 209.3$ GeV, $\quad m_{H_2^+} = 242.1$ GeV}\\
\multicolumn{4}{|c|}{$|\epsilon_K|= 2.218\times 10^{-3}$, $\quad \Delta m_K = 2.822 \times 10^{-15}$ GeV}\\ 
\multicolumn{4}{|c|}{$\Delta m_{B_d} = 3.841\times 10^{-13}$ GeV, $\quad \Delta m_{B_s} = 1.225\times 10^{-11}$ GeV}\\
\hline
\end{tabular}
\caption{\label{Tab: BP} Two benchmark points for our model implementation in cases $(B_1, B_1)$ and $(B_1,B_3)$
and their resulting mass spectra and predictions for the flavor observables.}
\end{center}
\end{table}
%


\section{Discussion and conclusions}
\label{section:discussion}

In this paper, we undertook the first step in the phenomenological investigation of a unique variant of 3HDM: the simplest model
which realizes the $CP$ symmetry of order 4 (CP4) without any accidental symmetry. The single symmetry CP4 strongly 
constrains both the scalar and Yukawa sectors, but it is nevertheless capable of accurately fitting all quark masses 
and mixing parameters, including $CP$ violation, as well as the SM-like Higgs boson properties.

When developing the model, we obtained convenient sufficient conditions for the stability and boundedness from below
for the potential, analytic expressions for the minima and scalar mass matrices, and the condition for alignment in the scalar sector.
We also established all forms of the CP4-invariant Yukawa Lagrangian with four distinct Yukawa textures in the up and down quark sectors.

The spontaneous breaking of the CP4 symmetry induces proper splittings in the fermion mass spectra consistent with observations. 
In the scalar alignment limit, the SM-like Higgs boson $H_{125}$, which we assumed to be the lightest scalar, was shown to conform 
to the experimental constraints on the FCNCs and the electroweak precision observables. The heavier Higgs partners do induce 
tree-level FCNC effects, but, thanks to a destructive interference among them, the resulting deviations from the SM values of kaon 
and $B$-meson mixing and CPV are kept under control.

To make the analysis quantitative, we have set up a conservative scan over the parameter space of the model
adopting the most relevant theoretical and phenomenological constraints from the scalar and Yukawa sectors.
Our $\chi^2$ fit has revealed viable parameter space regions for several distinct scenarios for the combinations 
of up and down Yukawa coupling matrices and showed a promising potential to rule out some of them.
The benchmark points found in this analysis could be very useful for follow-up in-depth phenomenological 
explorations of the CP4 3HDM in the future.

In this exploratory study we did not address many specific issues which merit close examination.
These directions of research include:
\begin{itemize}
\item
checking the compatibility of the points which pass meson oscillation constraints (red dots in the above plots)
with direct searches for new neutral and heavy Higgs scalars at the LHC.
Given that the masses of extra Higgses are rather low, one may suspect
that these points are already ruled out by the negative results of the heavy Higgs boson searches.
This is not the case since, in the scalar alignment regime, all extra Higgses decouple
from the gauge bosons. In addition, $t\bar t H_i$ coupling---and therefore the $gg$-fusion 
production cross section---may well be suppressed with respect to the SM, 
but even if it is not, the extra Higgses can easily avoid the main LHC searches which focus on the $ZZ/WW$
or $\gamma\gamma$ final states. One needs to look into the $q\bar q$ decay channels,
especially into flavor-violating ones, to see their manifestation.
In any case, a dedicated study is required to check what percentage of the points we have found
are ruled out by the direct LHC searches.
\item
checking whether the light charged Higgs bosons can satisfy $B$-decay constraints, especially $b\to s\gamma$.
It is known that, within 2HDM type-II, this decay places a very stringent lower bound on
the charged Higgs mass $m_{H^+} \gsim 580$ GeV, \cite{Misiak:2017bgg}. 
In our case, there are {\em two} charged Higgs bosons which may destructively interfere
due to the CP4-driven structural relations between the Yukawa matrices for $\Phi_2$ and $\Phi_3$.
We expect that, at least for some of the scan points, this interference will 
significantly reduce the combined charged Higgs contribution to $b \to s\gamma$
and relax the constraints.
\item
checking the magnitude of the electric dipole moments (EDMs) 
induced within CP4 3HDM and its compatibility with the negative
results of the electron and neutron EMD searches, see e.g. \cite{Jung:2013hka,Inoue:2014nva,Cheung:2014oaa} for summaries
of the experimental results and for analyses of EDMs within multi-Higgs models. 
The strongest constraints are placed by the electron EDM limits, but it can be easily avoided
by assuming that CP4 does not extend to the lepton sector.
In the quark sector, the $CP$-violating couplings are indeed present
and may enhance the neutron EDM with respect to the Standard Model.
Here we only want to give two remarks.
First, due to the scalar alignment, the SM-like Higgs boson exchanges by themselves
do not generate EDMs. Such contributions can only arise from exchanges of extra neutral Higgses.
The corresponding diagrams involve flavor-diagonal Yukawa couplings of $H_i$,
which may be suppressed in CP4 3HDM. In addition, the contribution of
the four extra scalars may exhibit a similar destructive interference driven by the structural properties 
of CP4-symmetric Yukawa sector. This issue certainly requires a dedicated study.
\item
investigating how close the CP4 3HDM can approach the SM and which observables exhibit the strongest deviations.
In multi-Higgs models such as 2HDM, it is customary to define the SM-like limit of the theory.
In CP4 3HDM, it is not guaranteed that this limit exists at all, again due to the structural 
relations among various parameters imposed by CP4 symmetry.
In our scan, the SM-limit was definitely absence due to the scalar alignment limit condition we imposed,
$m_{11}^2=m_{22}^2$. It remains an open question if such limit exists if $m_{22}^2$ is taken to be negative and large.
\item
checking whether the CP4 3HDM can alleviate any of the several tensions persistently observed
over the last few years in various $B$-meson decays;
\item
checking whether the particular variants of the CP4 3HDM such as the Yukawa combination $(A,A)$ or the presence
of new light scalars, which we briefly mentioned in the text, are ruled out or still compatible with the data;
\item
incorporating the charged leptons and verifying that the experimental bounds on lepton flavor violating processes
can be satisfied; 
\item
building neutrino mass models based on CP4 symmetry. A first step in this direction was take in 
\cite{Ivanov:2017bdx}.
\item
verifying if the CP4 3HDM can say something useful about the strong $CP$ problem.
\end{itemize}

The key message of this work is that, unlike in many other multi-Higgs models, a single symmetry requirement strongly shapes 
the model and brings us very far without falling into a direct conflict with experiment. It is true that on the way we had to stick 
to the exact scalar alignment, whose condition $m_{11}^2 = m_{22}^2$ is not symmetry-protected. However, it may turn out 
that the CP4 3HDM emerges as the lower energy limit of yet another model with a higher symmetry. In this case, $m_{11}^2 = m_{22}^2$ 
would arise naturally from a higher symmetry requirement at a certain energy scale, and then, when running down to the electroweak scale,
the parameters would not deviate much from the alignment condition. Again, verifying that this construction is viable requires 
a dedicated study.

Finally, one can imagine multi-Higgs models with $CP$ symmetries of even higher order: CP8, CP16, etc. These will require more than 
three Higgs doublets since the full classification of symmetries possible within the 3HDM 
is complete \cite{abelian,Ivanov:2012ry,Ivanov:2012fp} and it does not include such symmetries. 
Whether these can be extended to the Yukawa sector in a satisfactory manner remains to be seen.

In short, models based on a single higher-order $CP$ symmetry arise as an attractive minimalistic framework, with many phenomena 
yet to be explored.

\bigskip

{\bf Acknowledgments}
Useful discussions with Miguel Nebot are gratefully acknowledged. R.P. is partially supported by the Swedish Research Council,
contract number 621-2013-428 and by CONICYT grant PIA ACT1406. H.S.  has  received  funding  from  the  European  Re-
search Council (ERC) under the European Union’s Horizon 2020 research and innovation programme (grant agreement No 668679). The work of I.P.I. was supported by the Portuguese
\textit{Fun\-da\-\c{c}\~{a}o para a Ci\^{e}ncia e a Tecnologia} (FCT) through the Investigator contract IF/00989/2014/CP1214/CT0004
under the IF2014 Program and in part by contracts UID/FIS/00777/2013 and CERN/FIS-NUC/0010/2015,
which are partially funded through POCTI, COMPETE, QREN, and the European Union. P.F. and I.P.I. also acknowledge the support 
from National Science Center, Poland, via the project Harmonia (UMO-2015/18/M/ST2/00518).
The support from CONACYT project CB-2015-01/257655 (M\'exico) is also acknowledged by E.J. 

\appendix

\section{Simplifying the potential}\label{appendix-simplifying}

The phase-sensitive part of the CP4 3HDM potential $V_1$ first presented in \cite{abelian}
was of the form 
\be
V_1 =\lambda_5 (\phi_3^\dagger\phi_1)(\phi_2^\dagger\phi_1)
+{\lambda_6 \over 2} (\phi_2^\dagger\phi_1)^2 + {\lambda_7 \over 2} (\phi_1^\dagger\phi_3)^2 +
\lambda_8(\phi_2^\dagger \phi_3)^2 + \lambda_9(\phi_2^\dagger\phi_3)\left(\phi_2^\dagger\phi_2-\phi_3^\dagger\phi_3\right) + h.c.,
\label{V1initial}
\ee
where all coefficients could be complex.
Invariance under CP4 given by Eq.~(\ref{J-def}) immediately requires that $\lambda_5$ be real, 
$|\lambda_6|=|\lambda_7|$, and $\arg\lambda_6 - \arg\lambda_7 = \pi$.
This brings us to the for of $V_1$ analyzed in \cite{Ivanov:2015mwl,Aranda:2016qmp}:
\be
V_1 = \lambda_5 (\phi_3^\dagger\phi_1)(\phi_2^\dagger\phi_1)
+ {\lambda_6 \over 2} \left[(\phi_2^\dagger\phi_1)^2 - (\phi_1^\dagger\phi_3)^2\right] +
\lambda_8(\phi_2^\dagger \phi_3)^2 + \lambda_9(\phi_2^\dagger\phi_3)\left(\phi_2^\dagger\phi_2-\phi_3^\dagger\phi_3\right) + h.c.
\label{V1b}
\ee
Using rephasing of $\phi_2$ and $\phi_3$, 
both $\lambda_{5}$ and $\lambda_6$ can be made real, but then $\lambda_{8}, \lambda_9$ are in general complex.
It turns out that this form of the CP4 3HDM potential is not the simplest one.
Indeed, there exists a residual freedom of basis changes $\phi_i \mapsto \phi_i' = R_{ij}\phi_j$,
which leaves invariant the symmetry transformation $J$ if $RXR^T = X$.
This condition enforces $R$ to be of the block-diagonal form with $R_{11} = 1$ and
the $2\times 2$ block being a generic $Sp(2,\mathbb{C})$ transformation.
Absorbing the rephasings into redefinition of $\phi$'s, one gets one extra transformation freedom:
an orthogonal rotation between doublets $\phi_2$ and $\phi_3$.

This rotation {\em reparametrizes} the general CP4 3HDM potential.
The potential remains of the same form but new parameters are expressed via the old parameters
in a non-trivial way. In particular, it induces an orthogonal transformation between $\lambda_5$ and $\lambda_6$
with twice the $(\phi_2, \phi_3)$ rotation angle. This is so because, in group-theoretic terms, there exist
two bilinear combinations which pick up $i$ under CP4: $\phi_1^\dagger \phi_2 + \phi_3^\dagger \phi_1$
and $\phi_1^\dagger \phi_3 - \phi_2^\dagger \phi_1$, and it is the product of these two bilinears with their
conjugates that is encoded in the $\lambda_5$ and $\lambda_6$ terms.

Therefore, starting from $V_{1}$ given by Eq.~\eqref{V1b},
one can always find a basis in which $\lambda_5 = 0$ or $\lambda_6= 0$.
In the present work, we choose the latter option and eliminate $\lambda_6$.
We could have used the former one; in fact, that choice would be preferred for investigation
of 3HDM with unbroken CP4. We stress that when setting one of these parameters to zero
we do not lose any generality.

\section{Boundedness from below}\label{appendix-BFB-conditions}

In the 2HDM, the exact necessary and sufficient boundedness-from-below (BFB) conditions are known
explicitly for the most general renormalizable scalar potential \cite{Ivanov:2006yq}.
Beyond two doublets, this general result is still unknown.
Although in certain simplified or symmetric cases one can establish them,
we could not find such necessary and sufficient conditions for the CP4 3HDM.

For the purpose of a numerical scan, we limited ourselves to a set of {\em sufficient conditions},
which constrain the parameters somewhat stronger than needed but which guarantee
that the models selected have stable potentials. Here we outline how these conditions were derived.

Let us denote
\be
r_i \equiv \phi_i^\dagger \phi_i \ge 0\,,\quad
z_{ij} \equiv (\phi_i^\dagger \phi_i) (\phi_j^\dagger \phi_j) - (\phi_i^\dagger \phi_j) (\phi_j^\dagger \phi_i) \ge 0\,. \label{ri-zij}
\ee
All three quantities $z_{ij}$ are independent in 3HDM (they become dependent for $N>3$ doublets).
Then, the phase-insensitive potential can be written as
\be
V_0 = \lambda_1 r_1^2 + \lambda_{2}(r_2^2 + r_3^2) + \lambda_{34} r_1(r_2+r_3) + \lambda'_{34}r_2r_3
- \lambda_4 (z_{12} + z_{13}) - \lambda'_4 z_{23}\,,\label{V0-rz}
\ee
where, as usual, $\lambda_{34} \equiv \lambda_3 + \lambda_4$ and $\lambda'_{34} \equiv \lambda'_3 + \lambda'_4$.
If this were the full potential, we could have established the exact necessary and sufficient BFB conditions
via the so-called copositivity conditions:
the conditions that the real symmetric quadratic form in $r_i$ is positive definite in the positive orthant $r_i \ge 0$.
Applications of copositivity conditions to various Higgs potentials can be found in \cite{Kannike:2012pe,Chakrabortty:2013mha}.
In the simplest case of a $2\times 2$ matrix $a_{ij}$ these conditions are
\be
\quad a_{11}> 0\,,\quad a_{22}>0\,,
\quad \tilde a_{12} \equiv a_{12} + \sqrt{a_{11}a_{22}} > 0\,. \label{copositivity2x2}
\ee
For a $3\times 3$ matrix, the copositivity conditions are \cite{Kannike:2012pe}
\bea
&&  a_{11}> 0\,,\quad a_{22}>0\,, \quad a_{33}>0\,, \quad
\tilde a_{12} > 0\,, \quad \tilde a_{13} > 0\,, \quad \tilde a_{23} > 0\,, \nonumber\\
&& \sqrt{a_{11}a_{22}a_{33}} + a_{12}\sqrt{a_{33}} + a_{13}\sqrt{a_{22}} + a_{23}\sqrt{a_{11}}
+ \sqrt{2 \tilde a_{12} \tilde a_{13} \tilde a_{23}} >0\,.\label{copositivity=3x3}
\eea
In addition, since we want the minimum to be neutral, it is sufficient to require
that $\lambda_4 < 0$ and $\lambda'_4 < 0$.

In our case, the potential also contains phase-sensitive terms $V_1$ which depend not only on $r_i$
but also on the relative phases of the fields. We can rewrite $V_1$ as
\be
V_1 = - 2 |\lambda_5| r_1\sqrt{r_2r_3} \cos\psi_5 - 2|\lambda_8| r_2 r_3 \cos\psi_8 - 2|\lambda_9| \sqrt{r_2r_3}(r_2-r_3) \cos\psi_9 \,.
\label{V1-ri}
\ee
Here, the factors $\cos\psi_i$ take into account the relative phases between the fields as well as the phases of the coefficients.
Since $|\cos\psi_i| \le 1$ and $\sqrt{r_2r_3} \le (r_2+r_3)/2$, we can establish a lower bound $V_1 > V'_1$, where
\be
V'_1 = - |\lambda_5| r_1(r_2+r_3) - 2|\lambda_8| r_2 r_3 - |\lambda_9| (r_2^2-r_3^2)\,,\label{V1-lower}
\ee
where without loss of generality we replaced $|r_2^2-r_3^2|$ with just $r_2^2-r_3^2$.
Since $V_0 + V_1 > V_0 + V'_1 = a_{ij} r_i r_j$, we apply the copositivity constraints on $a_{ij}$, where
\bea
&&a_{11} = \lambda_1\,, \quad a_{22} = \lambda_2 - |\lambda_9|\,, \quad a_{33} = \lambda_2+ |\lambda_9|\,, \nonumber\\
&& a_{12} = a_{13} = {1\over 2}\lambda_{34} - {1\over 2}|\lambda_5|\,,\quad
a_{23} = {1\over 2}\lambda'_{34} - |\lambda_8|\,. \label{matrixA}
\eea
These conditions will automatically imply that the original potential is bounded from below.
They may be overly restrictive and can perhaps be improved, but they are sufficient
for the purposes of our numerical scan.

\section{Three scalar bases}\label{appendix-three-bases}

Here we define the three bases for the scalar fields, appropriate
for the discussion of the scalar alignment: the original basis,
one specific choice of the Higgs basis, and the physical scalar basis.

In the original basis the neutral fields are defined as $\phi_i^0 = (h_i + i a_i)/\sqrt{2}$.
Their vevs are parametrized as \eqref{vevs2}.
In the 6D real field space $\varphi_a = (h_1, h_2, h_3, a_1, a_2, a_3)$, the vev direction
is parametrized as
\be
\lr{\varphi_a} = (c_\psi, s_\psi c_\beta c_\gamma, s_\psi s_\beta c_\gamma, 0, s_\psi c_\beta s_\gamma, -s_\psi s_\beta s_\gamma)\,.
\ee

A {\em Higgs basis} is defined as a basis in which only the first doublet gets the vevs.
The neutral complex fields are written in this basis as
\be
\Phi_i = \triplet{\Phi_1}{\Phi_2}{\Phi_3} \equiv {1\over\sqrt{2}}\triplet{\rho_1 + i G^0}{\rho_2 + i \eta_2}{\rho_3 + i \eta_3}\,,
\quad \lr{\Phi_i} = \triplet{v}{0}{0}\,.\label{Higgs-c}
\ee
Here, we already took into account that $\eta_1 \equiv G^0$ is the neutral Goldstone boson.
In this basis, we define the 6D real field space as
\be
\Phi_a = (G^0, \rho_1, \rho_2, \rho_3, \eta_2, \eta_3)\,, \quad
\lr{\Phi_a} = (0, v, 0, 0, 0, 0)\,.\label{Higgs-r}
\ee
Finally, the {\em physical scalar basis}, which is defined as the basis in which the neutral mass matrix ${\cal M}_n$ is diagonal,
must be written via real fields, which we label as
\be
H_a = (G^0, H_{125}, H_3, H_4, H_5, H_6)\,.\label{physical}
\ee
In 3HDM, the Higgs basis is not uniquely defined: one can always perform a unitary transformation
between $\Phi_2$ and $\Phi_3$ in Eq.~\eqref{Higgs-c} preserving the definition of the Higgs basis.
Here, we use the following convenient and traditional choice:
\be
\Phi_i = {\cal P}_{ij} \phi_j\,, \qquad
\triplet{\Phi_1}{\Phi_2}{\Phi_3} =
\mmmatrix{c_\beta}{s_\beta c_\psi}{s_\beta s_\psi}{0}{-s_\psi}{c_\psi}{s_\beta}{-c_\beta c_\psi}{-c_\beta s_\psi}
\triplet{\phi_1}{\phi_2 e^{-i\gamma}}{\phi_3 e^{i\gamma}}\,.
\ee
In terms of real fields, we get $\Phi_a = P_{ab} \varphi_b$, where
\be
P_{ab} = \left(\begin{array}{cccccc}
0 & - s_\beta c_\psi s_\gamma & s_\beta s_\psi s_\gamma & c_\beta & s_\beta c_\psi c_\gamma & s_\beta s_\psi c_\gamma \\[1mm]
c_\beta & s_\beta c_\psi c_\gamma & s_\beta s_\psi c_\gamma & 0 & s_\beta c_\psi s_\gamma & -s_\beta s_\psi s_\gamma \\[1mm]
0 & - s_\psi c_\gamma & c_\psi c_\gamma & 0 & - s_\psi s_\gamma & - c_\psi s_\gamma \\[1mm]
s_\beta & -c_\beta c_\psi c_\gamma & -c_\beta s_\psi c_\gamma & 0 & -c_\beta c_\psi s_\gamma & c_\beta s_\psi s_\gamma \\[1mm]
0 & s_\psi s_\gamma & c_\psi s_\gamma & 0 & - s_\psi c_\gamma & c_\psi c_\gamma \\[1mm]
0 & c_\beta c_\psi s_\gamma & -c_\beta s_\psi s_\gamma & s_\beta & -c_\beta c_\psi c_\gamma & -c_\beta s_\psi c_\gamma
\end{array}\right)\,.\label{rotation-matrix-P}
\ee
The physical scalar basis is linked with the Higgs basis and the original basis via matrices $Q$ and $R$, respectively:
\be\label{falvor-mass-R}
H_a = Q_{ab} \Phi_b = R_{ac} \varphi_c\,,\quad  R_{ac} = Q_{ab} P_{bc}\,.
\ee
Notice that since the Goldstone does not mix with the physical Higgses, the matrix $Q$ is of block-diagonal form:
\be
Q_{ab} = \mmatrix{1}{0}{0}{Q_{5\times 5}}\,.
\ee

\end{document}